\pdfoutput=1
\documentclass[aps,prd, amsmath,amssymb,superscriptaddress,showkeys,nofootinbib,reprint,notitlepage,preprintnumbers,onecolumn]{revtex4-1}
\usepackage{color}
\usepackage{aas_macros}
\usepackage{hyperref,breakurl}
\usepackage[usenames,dvipsnames]{xcolor}
\usepackage{dcolumn}
\usepackage{bm}
\usepackage{hyperref}
\usepackage{array}
\usepackage{dcolumn}
\usepackage{enumerate}
\usepackage{nicefrac}
\usepackage{graphicx}
\usepackage{natbib}
\usepackage{siunitx}

\newcommand{\comment}[1]{} % Suppress a large block

\newcommand{\Pl}{P_{\rm lin}} 
\newcommand{\Om}{\Omega_{\rm m}} 

\allowdisplaybreaks

\newcommand{\ben}{\begin{enumerate}}
\newcommand{\bi}{\begin{itemize}}
\newcommand{\ei}{\end{itemize}}
\newcommand{\een}{\end{enumerate}}

\newcommand{\sig}{\sigma_8}

\newcommand{\lcdm}{$\Lambda$CDM}

\newcommand\fig[1]{Figure~\ref{#1}}

\DeclareSIUnit \h {\mbox{$h$}}
\DeclareSIUnit \degsq {\mbox{$\rm{deg}^2$}}

\DeclareSIUnit \parsec {pc}
\DeclareSIUnit \parsec {pc}
\DeclareSIUnit \megaparsec {Mpc}

\begin{document}

%%%%%%%%%%%%%%%%%%%%%%%%%%%%%%%%%%%%%%%%%%%%%%%%%%%%%%%

\title{Beyond linear galaxy alignments}
\author{Jonathan A.~Blazek}
\email{blazek@berkeley.edu}
\affiliation{Institute of Physics, Laboratory of Astrophysics, \'Ecole Polytechnique F\'ed\'erale de Lausanne (EPFL), Observatoire de Sauverny, 1290 Versoix, Switzerland}
\affiliation{Center for Cosmology and AstroParticle Physics, Department of Physics, The Ohio State University, 191 West Woodruff Avenue, Columbus, Ohio 43210, USA}
\author{Niall MacCrann}
\author{M.~A.~Troxel}
\author{Xiao Fang}
\affiliation{Center for Cosmology and AstroParticle Physics, Department of Physics, The Ohio State University, 191 West Woodruff Avenue, Columbus, Ohio 43210, USA}

\date{\today}

\begin{abstract}

Galaxy intrinsic alignments (IA) are a critical uncertainty for current and future weak lensing measurements. We describe a perturbative expansion of IA, analogous to the treatment of galaxy biasing. From an astrophysical perspective, this model includes the expected large-scale alignment mechanisms for galaxies that are pressure-supported (tidal alignment) and rotation-supported (tidal torquing) as well as the cross-correlation between the two. Alternatively, this expansion can be viewed as an effective model capturing all relevant effects up to the given order. We include terms up to second order in the density and tidal fields and calculate the resulting IA contributions to two-point statistics at one-loop order. For fiducial amplitudes of the IA parameters, we find the quadratic alignment and linear-quadratic cross terms can contribute order-unity corrections to the total intrinsic alignment signal at $k\sim0.1\,h^{-1}{\rm Mpc}$, depending on the source redshift distribution. These contributions can lead to significant biases on inferred cosmological parameters in Stage IV photometric weak lensing surveys. We perform forecasts for an LSST-like survey, finding that use of the standard ``NLA'' model for intrinsic alignments cannot remove these large parameter biases, even when allowing for a more general redshift dependence. The model presented here will allow for more accurate and flexible IA treatment in weak lensing and combined probes analyses, and an implementation is made available as part of the public {\tt FAST-PT} code. The model also provides a more advanced framework for understanding the underlying IA processes and their relationship to fundamental physics. 

\end{abstract}

%\pacs{Valid PACS appear here}
%\keywords{gravitational lensing: weak; dark matter; dark energy; cosmology: theory; cosmology: observations; cosmological parameters}

\maketitle

%%%%%%%%%%%%%%%%
\section{Introduction}
%%%%%%%%%%%%%%%%

State-of-the-art weak lensing surveys will measure correlations between the shapes of galaxies to unprecedented levels of precision. These measurements provide precise constraints on cosmological models and contribute to key tests of gravity on large scales. However, these correlations are sourced not only by lensing of background galaxies, but also by the intrinsic shapes and alignments of galaxies and the non-negligible cross-correlation between these effects. These ``intrinsic alignments'' (IA) are an interesting physical phenomenon that depend on the formation and evolution of galaxies as well as the relationship between galaxies, their surrounding halos, and the underlying large-scale structure (for recent reviews on IA, see \cite{troxel15rev,joachimi15rev}).

The processes underlying the correlated alignments of galaxies remain uncertain, and it is likely that different mechanisms may be relevant for galaxies with different kinematic properties. Most notably, galaxies whose orientations are set by angular momentum and those with primarily pressure support are likely to exhibit different alignment behavior. These kinematic types are typically classified through morphological properties as ``spirals'' and ``ellipticals,'' respectively, or even using color (e.g.\ ``blue'' and ``red'') as a proxy. However, it may be that a more subtle distinction is necessary, including e.g.\ the difference between rapidly and slowly rotating ellipticals \cite{kormendy09}.

For weak lensing measurements in current and upcoming photometric surveys,\footnote{e.g.\ Dark Energy Survey (DES), https://www.darkenergysurvey.org; \\Kilo Degree Survey (KiDS), http://kids.strw.leidenuniv.nl/; \\Hyper Suprime-Cam(HSC), http://www.subarutelescope.org/Projects/HSC/; \\Euclid, http://sci.esa.int/science-e/www/area/index.cfm?fareaid=102; \\Large Synoptic Survey Telescope (LSST), http://www.lsst.org; \\and Wide-Field Infrared Survey Telescope (WFIRST), http://wfirst.gsfc.nasa.gov/} including an appropriate model for IA is necessary to avoid biased inference of cosmological parameters \cite{joachimi10,krause16}. To date, most analyses (e.g.\ \cite{heymans13,DES15,hildebrandt17}) have used some version of the ``tidal alignment'' model \cite{hirata04,bridle07,blazek11}. However, this theory has only been definitively shown to describe massive elliptical galaxies on large scales \cite{hirata07,mandelbaum11,joachimi11,singh15,okumura09b,west17}. Spirals, and less luminous ellipticals, dominate the populations of galaxies found in typical weak lensing measurements, which are also sensitive to correlations on smaller scales. The most recent observations suggest that these typical lensing sources may also exhibit IA \cite{troxel17arxiv,des_keypaper_17arxiv,vanuitert17arxiv}.
Improved understanding, measurement, and mitigation of IA are critical for the success of future lensing experiments, which will achieve exquisite statistical precision and thus will be dominated by systematic uncertainties.

One potential approach to simplify the modeling of IA in weak lensing measurements would be to split the source population (e.g.,\ by color or morphological classification) into groups in which a single IA mechanism is expected to dominate. However, such source splits may remove statistical power or increase the complexity of the analysis (e.g.\ if shear calibration must be separately performed on each sub-sample). Moreover, as discussed below, it is likely that multiple alignment mechanisms apply to a given source sample (analogous to multiple contributions to galaxy bias). It is thus critical to develop a general model for IA, including all relevant contributions and cross-correlations. In this paper, we present a mixed model for IA, including all tidal alignment and tidal torquing effects up to second order. This model is a natural extension to the nonlinear tidal alignment model described in \cite{blazek15} and incorporates second-order contributions from the tidal torquing model \cite{catelan01,crittenden01,mackey02,hirata04}, including mixed terms between galaxies with alignments sourced by different mechanisms.

This perturbative approach is inspired by analogous work in galaxy bias (e.g.\ \cite{mcdonald09b,desjacques16arxiv}). The underlying principle is that all potential contributions at a given order, consistent with the required symmetries of the observable, are included. Each term carries an amplitude parameter which can receive contributions from higher-order correlations through ``renormalization.'' In this manner, the effect of small-scale physics on correlations at larger scales (where a perturbative expansion is sensible) are naturally included. We believe this approach will form a foundation for a more rigorous treatment of IA and will benefit from the significant insights available from studies of galaxy biasing. Moreover, unifying the treatment of biasing and IA will benefit future analyses that rely on multiple probes (e.g.\ galaxy clustering and weak lensing) to measure both cosmological parameters and astrophysical effects. Finally, in addition to their importance in weak lensing measurements, IA can provide a powerful probe of both astrophysics and new fundamental physics (e.g.\ \cite{schmidt15}). The approach outlined here provides a useful framework for including the potential IA signatures of these effects. The model described here has been implemented in the publicly available {\tt FAST-PT} code \cite{mcewen16,fang17} and has been applied to the current state-of-the-art cosmic shear analysis \cite{troxel17arxiv}.

The structure of the paper is as follows. In Sec.~\ref{sec:theory}, we summarize the background concepts and describe the IA expansion in terms of cosmological fields. In Sec.~\ref{sec:corr}, we calculate the relevant correlations for two-point weak lensing observables and discuss a number of related details, including renormalization of the IA parameters and scaling the overall amplitudes. Sec.~\ref{sec:implementation} describes how we implement the model and presents the resulting contributions to cosmic shear statistics. We also present a forecast for the impact of this IA model on an LSST-like survey. We conclude in Sec.~\ref{sec:conc}. In an appendix, we present the analytic forms of the IA power spectra as well as the decomposition of the terms into the basis used for {\tt FAST-PT} evaluation.
Where relevant, we assume a flat $\Lambda$CDM cosmology with $\Omega_m=0.315$, $\sigma_8=0.82$, $h=0.67$, $\Omega_b=0.044$, $n_s=0.96$ and that a single massive neutrino eigenstate provides a neutrino density $\Omega_{\nu}h^2=6.5\times10^{-4}$.

%%%%%%%%%%%%%%%%
\section{Perturbative expansion for IA}
\label{sec:theory}
%%%%%%%%%%%%%%%%

Our goal is to construct a perturbative IA model that includes both tidal alignment (linear in the tidal field) and tidal torquing (quadratic in the tidal field) terms, as well as their cross-correlations. In the following, we motivate this approach and provide a brief summary of quantities that will be directly useful for our calculations.

\subsection{Preliminary definitions and conventions}

We begin by considering perturbative expansions of galaxy bias, where the relevant symmetry is that all contributions must be scalars. Following the notation of \cite{baldauf12}, we can write a local bias model complete to second order in the matter density and tidal fields, ignoring higher derivative terms and functions of the velocity divergence $\theta$ which enter at higher order:
\begin{align}
\label{eq:biasexp}
\delta_g(\mathbf{x}) = b_1 \delta(\mathbf{x}) + \frac{b_2}{2} \left(\delta(\mathbf{x})^2 - \langle \delta^2\rangle \right) + \frac{b_s}{2} \left(s(\mathbf{x})^2 - \langle s^2\rangle \right) + \cdots~,
\end{align}
where $\delta$ and $s$ are the (nonlinear) density and tidal fields, which can be expanded in terms of the linear density field and relevant gravity kernels in standard perturbation theory (SPT; e.g.\ \cite{bernardeau02}):
\begin{align}
\delta = \delta^{(1)}+ \delta^{(2)}+ \delta^{(3)}+\cdots~.
\end{align}
It is convenient to work in Fourier space, where we define power spectra in terms of the ensemble average over the Fourier space fields:
\begin{align}
\langle A(\mathbf{k}) B(\mathbf{k'})\rangle = (2\pi)^3\delta_D^{(3)}(\mathbf{k}+\mathbf{k'})P_{AB}(k)~.
\end{align}
The linear density field is $\delta^{(1)}$. The second-order contribution is:
\begin{align}
\delta^{(2)}(\mathbf{k}) &= \int \frac{ d^3 \mathbf{k}_1}{(2 \pi)^3} F_2( \mathbf{k}_1, \mathbf{k}_2) \delta^{(1)}(\mathbf{k}_1) \delta^{(1)}(\mathbf{k}_2)~,
\end{align} 
where
$\mathbf{k}_2 \equiv \mathbf{k}-\mathbf{k}_1$, $\mu_{12} \equiv \hat{k}_1\cdot\hat{k}_2$, and the second-order density kernel is
\begin{align}
F_2(\mathbf{k}_1, \mathbf{k}_2) = \frac{5}{7} + \frac{1}{2}\mu_{12}\left( \frac{k_1}{k_2}+\frac{k_2}{k_1} \right) + \frac27\mu_{12}^2~.
\end{align}
We define the normalized Fourier-space tidal tensor $s_{ij}$:
\begin{align}
s_{ij}(\mathbf{k}) &=  \left(\hat{k}_{i} \hat{k}_{j} - \frac{1}{3}\delta_{ij}\right)\delta(k) \equiv \hat{S}_{ij}\left[\delta(k)\right]~.
\end{align}
The squared tidal tensor is then
\begin{align} 
s^2(\mathbf{k})= \int \frac{d ^3 \mathbf{k}_1 }{(2 \pi)^3} S_2( \mathbf{k}_1, \mathbf{k}_2 ) \delta(\mathbf{k}_1) \delta(\mathbf{k}_2) ~, 
\end{align}
where $S_2(\mathbf{k}_1, \mathbf{k}_2) = \mu_{12}^2 - \frac{1}{3}$.
More complicated bias expansions are possible, for instance including the density and velocity of the relative baryon-dark matter fluid \cite{blazek15, schmidt16} as well as higher-order derivative contributions -- see \cite{desjacques16arxiv} for a detailed review.

\subsection{Application to IA}
We now apply these techniques to develop a general perturbative expansion for IA. The basic procedure is the same as with galaxy bias, except that we now use (local) functions of the cosmological fields with the correct symmetry for galaxy shapes, namely traceless, symmetric tensors. We begin by expanding in 3d quantities. In the next sub-section, we project these quantities onto the plane of the sky to obtain expressions for the projected shapes, which will have spin-2 symmetry.\footnote{Since shape measurement is not a linear process, the measured 2d shape is not necessarily the same as the projected 3d shape. We leave this subtlety for future consideration.} We will then decompose these projected shapes into $E$- and $B$-mode components before calculating the relevant correlations. This approach is only exact in the flat-sky approximation -- for a more rigorous treatment of this projection in the full-sky regime, see \cite{schmidt15}.

Expanding up to second-order in the linear density field, and as before considering contributions from only the total matter (rather than the DM-baryon relative fluid), we have:
\begin{align}
\label{eq:IAexp}
\gamma^I_{ij}(\mathbf{x}) = C_1 s_{ij} + C_2 \left(s_{ik} s_{kj} - \frac{1}{3} \delta_{ij} s^2\right) + C_{1\delta} \left( \delta s_{ij} \right) + C_{t} t_{ij} + \cdots ~,
\end{align}
where all fields are evaluated at $\mathbf{x}$ and summation over repeated indices is implied. The $C_i$ parameters are analogous to galaxy bias parameters, capturing the effective strength of each term, including the contributions from small-scale physics (see also \cite{schmidt15} for a similar treatment of IA in the context of non-Gaussianity). In this expansion, we have included the tensor $t_{ij} = \hat{S}_{ij} [\theta - \delta]$, which involves the velocity shear (see \cite{mcdonald09b}). Due to spin-2 symmetry of galaxy shapes, the fields $s_{ij}$ and $t_{ij}$ both contribute to $\gamma^I_{ij}$ at lower order than in the case of the (scalar) galaxy density. The tidal field $s_{ij}$ provides the linear contribution, rather than appearing at second order, and $t_{ij}$ enters at second rather than third order. We note that the ``intrinsic shape'' of a galaxy is not a uniquely defined quantity and will depend on the shape measurement technique (see, e.g., \cite{singh16}), which will also impact the measured values of the $C_i$ parameters. The relationship between measured shapes and the underlying gravitational shear is similarly dependent on the particular measurement technique. Eq.~\ref{eq:IAexp} can thus be considered to describe either the intrinsic ``shapes'' of galaxies or the intrinsic ``shears'' (i.e.\ the quantity added to the gravitational shear when modeling the underlying signal). Note that we do not explicitly include stochastic shape contributions in this expansion, an issue discussed in Sec.~\ref{sec:shapenoise}.

We treat this model in Eulerian perturbation theory, evaluating all quantities at the observed position of the galaxy. When including the $t_{ij}$ term, the above expansion is complete to second order, neglecting higher-order derivatives of these terms, which are suppressed on large scales. The expansion can thus be consistently evaluated at either Eulerian or Lagrangian galaxy positions. The mapping between the two, reflecting galaxy advection, is captured by relationships between the terms in the complete expansion (see \cite{schmitz_inprep}, or for discussion in the context of galaxy biasing \cite{baldauf12,chan12,saito14}). Interestingly, $t_{ij}$ plays a similar role as its third-order analog in the galaxy biasing case ($b_{3,{\rm NL}}$, see \cite{saito14}) capturing nonlinear evolution of the tidal field (or, equivalently, dependence of intrinsic shapes on the tidal field at the Lagrangian position, or position of formation, rather than the Eulerian position). We will consider the $t_{ij}$ contribution to IA in separate work \cite{schmitz_inprep} and will neglect it in what follows. Thus, the correlations calculated below are not formally equivalent between Eulerian and Lagrangian treatments. Similarly, since we have not included contributions to the intrinsic shapes at third order in the linear density field, the two-point correlations are not complete at one-loop, i.e.\ $\mathcal{O}(\Pl^2)$. We leave these issues for future work.

A basic requirement for this expansion is that it includes the existing astrophysically-motivated models for IA, namely tidal alignment (linear) for pressure-supported galaxies and tidal torquing (quadratic) for galaxies dominated by angular momentum \cite{catelan01,hirata04}. Indeed, these models are captured by the $C_1$ and $C_2$ terms. However, as a perturbative expansion, the amplitude of these terms can now capture all relevant effects at this order, regardless of their astrophysical origin. Similarly, the $C_{1\delta}$ term can be motivated from the fact that the intrinsic shape field is measured only at the positions of galaxies, and the observable quantity is $\tilde{\gamma}^{I} = (1+\delta_g)\gamma^I$.\footnote{This is analogous to the fact that it is the galaxy momentum rather than the velocity field that is observable.} Linear galaxy biasing naturally produces an IA term corresponding to $C_{1\delta} = b_1 C_1$ (and similar for higher-order contributions, such as $C_{2\delta} = b_1 C_2$). However, as with the other terms, $C_{1\delta}$ can be thought of more generally as capturing any alignment physics with effective large-scale correlations that depend on $\delta s_{ij}$, and in this context can have a value different from the density-weighting prediction.

\subsection{Projection and shape components}

Equation~\ref{eq:IAexp} is expressed in terms of 3d quantities. However, we observe shapes projected onto the plane of the sky, which have two components:
\begin{align}
(\gamma_1,\gamma_2) = \gamma_0(\cos 2\phi,\sin 2\phi)~,
\end{align}
with the angle $\phi$ measured with respect to some fixed coordinate system. In configuration space measurements, this angle is typically measured with respect to the separation vector between galaxies, in which case these components are typically referred to as $(\gamma_+,\gamma_\times)$. In Fourier space, it is best to decompose into curl-free ($E$) and divergence-free ($B$) components, as is done in standard weak lensing and CMB polarization measurements \cite{kamionkowski98,mackey02}. Since this decomposition is coordinate independent, we are free to choose a convenient coordinate system for performing calculations. We put the $x-y$ plane on the sky and measure shape components with respect to the $x$-axis. In this case, we have:
\begin{align}
\label{eq:E_B_defined}
{\gamma}_E(\mathbf{k}) &= p(\hat{k})^{-1}\left[f_{E}(\hat{k}){\gamma}_{+}({\bf k}) + f_{B}(\hat{k}){\gamma}_{\times}({\bf k})\right] ~,\\
{\gamma}_B(\mathbf{k}) &= p(\hat{k})^{-1}\left[-f_{B}(\hat{k}){\gamma}_{+}({\bf k}) + f_{E}(\hat{k}){\gamma}_{\times}({\bf k})\right] ~,
\end{align}
where we have defined the angular operators:
\begin{align}
f_E(\hat{u}) &= \hat{u}_x^2 - \hat{u}_y^2 ~, ~~
f_B(\hat{u}) = 2\hat{u}_x \hat{u}_y ~,
\end{align}
as well as the projection operator $p(\hat{u}) = 1-\hat{u}_z^2 \equiv 1-\mu_u^2$, which removes the unobservable line-of-sight ellipticity. Because the $f_{(E,B)}$ operators already include the relevant projection, we include $p^{-1}$ to avoid projecting twice. In this coordinate system, the observable ellipticity components are:
\begin{align}
(\gamma_+,\gamma_{\times}) = (C_1 + C_{1 \delta} \delta) (s_{xx}-s_{yy}, 2s_{xy}) + \left(C_2 + C_{2 \delta}\delta\right) (s_{xk}s_{xk} - s_{yk}s_{yk},2s_{xk}s_{yk}) + \cdots ~,
\end{align}
where products of cosmological fields in configuration space are convolutions in Fourier space. We have included the third-order $C_{2 \delta}$ term, although as we see below it does not contribute to two-point correlations at one-loop order. We then have:
\begin{samepage}
\begin{align}
\gamma_{(E,B)} (\mathbf{k})&= C_1 f_{(E,B)}(\hat{k}) \delta(k) + C_{1\delta} \int \frac{d^3\mathbf{k_1}}{(2\pi)^3} f_{(E,B)}(\hat{k}_1) \delta(k_1) \delta(k_2) \\
&+ C_2 \int \frac{d^3\mathbf{k_1}}{(2\pi)^3} h_{(E,B)}(\hat{k}_1,\hat{k}_2) \delta(k_1) \delta(k_2)\notag \\
&+ C_{2\delta} \int \frac{d^3\mathbf{k_1}}{(2\pi)^3}  \frac{d^3\mathbf{k_2}}{(2\pi)^3} h_{(E,B)}(\hat{k}_1,\hat{k}_2) \delta(k_1) \delta(k_2) \delta(k_3)+ \cdots ~, \notag
\end{align}
where $\mathbf{k_1}+\mathbf{k_2} +\mathbf{k_3}= \mathbf{k}$, with $\mathbf{k_3}=0$ in all but the final term. We have defined the additional angular operators (e.g.\ \cite{hirata04}):
\end{samepage}
\begin{align}
h_E(\hat{u},\hat{v}) &= \hat{u}\cdot\hat{v}\left(\hat{u}_x \hat{v}_x-\hat{u}_y \hat{v}_y\right) - \frac{1}{3}\left( \hat{u}_x^2 +\hat{v}_x^2 - \hat{u}_y^2 -\hat{v}_y^2 \right) ~,  \\
h_B(\hat{u},\hat{v}) &= \hat{u}\cdot\hat{v}\left( \hat{u}_x \hat{v}_y + \hat{u}_y \hat{v}_x\right) -\frac{2}{3} \left(\hat{u}_x \hat{u}_y+\hat{v}_x \hat{v}_y\right) ~, 
\end{align}
where we have explicitly symmetrized the $h_B$ operator in its arguments ($h_E$ is naturally symmetric).

%%%%%%%%%%%%%%%%
\section{IA Correlations}
\label{sec:corr}
%%%%%%%%%%%%%%%%

In cosmic shear, where correlations of pairs of galaxy shapes are measured, there are two relevant IA contributions: intrinsic-intrinsic (``II''), in which the intrinsic galaxy shapes are correlated with each other due to physical proximity; and gravitational-intrinsic (``GI''), in which the same large-scale structure induces a lensing shear in one galaxy and influences the intrinsic shape of the other \cite{hirata04}. In total, when combined with the gravitational-gravitational term (``GG'', i.e.\ the standard lensing signal), the observed correlation between source bins $i$ and $j$ is given by:
\begin{align}
\label{eq:Pobs}
P_{ij}^{\rm obs} = P_{ij}^{\rm GG} + P_{ij}^{\rm GI} + P_{ij}^{\rm IG} + P_{ij}^{\rm II} ~.
\end{align}
Eq.~\ref{eq:Pobs} is typically considered for $E$-mode correlations. Beyond leading order, there can be $B$-mode contributions to both the gravitational and intrinsic shape contributions, although the lensing $B$-modes are suppressed on all but the smallest scales \cite{hirata04}. In the following, we consider $B$-modes from the II term only. Under parity, all $EB$ correlations must be zero.

While we group these calculations by the associated astrophysical model (i.e.\ tidal alignment and tidal torquing), the correlations can be combined for a general IA expansion (Sec.~\ref{sec:complete}).
In the following, we calculate the GI and II terms. We express the correlations with a single set of IA parameters $C_i$, which could be thought of as the parameters for a homogenous galaxy population or the effective average parameters for a mixed population (weighted by the fraction of each sub-population). The generalization to cross-correlations between sub-populations is straightforward. It is also straightforward to extend the model to include the gI term, the correlation between galaxy density and intrinsic shapes, which can impact galaxy-galaxy lensing and ``combined probe'' measurements (e.g.\ \cite{blazek12,des_keypaper_17arxiv}) and provides the most straightforward method for directly measuring IA. For linear galaxy biasing, the gI term is given by the GI correlations presented here, multiplied by the galaxy bias, while it will contain additional contributions due to nonlinear galaxy biasing (see \cite{blazek15} for the tidal alignment case). 

In this section, we derive the structure of the terms -- the full analytic form of each convolution integral is given in Appendix~\ref{app:kernels}. For simplicity, the redshift dependence has been suppressed. Because the terms are expansions in the linear power spectrum, this dependence is simple, with one-loop correlations scaling as $G(z)^4$, for linear growth factor $G(z)$. As noted above, the projection operator $p$ is implicitly included in our definitions of the $f$ and $h$ operators. While $p$ factors out in most correlations, it does not in cases where there are convolutions involving two projections (i.e.\ shape-shape correlations beyond linear order). The expressions are simplified in the Limber approximation, where only transverse modes ($\mu_k = 0 \rightarrow p(\hat{k}) = 1$) contribute. The Limber approximation is generally valid on scales of interest in weak lensing, since the lensing kernel and broad source redshift bins lead to a large projection length. In the appendix, we present the full expressions, which depend on $\mu_k$, but we use the Limber approximation when actually evaluating them. In the calculations below, we use the shorthand notation $f(\delta)$ or $h(\delta,\delta)$ to denote these angular operators acting on a density field (or two density fields, including the convolution, for $h$).%\footnote{See \cite{schmidt15} for an alternative treatment of this projection.}

\subsection{Tidal alignment}

The tidal alignment model at one-loop order, i.e.\ the $C_1$ and $C_{1\delta}$ terms, is presented in \cite{blazek15}. Here, we summarize the relevant contributions.

\subsubsection{GI correlation}
\label{sec:taGI}
The GI term is given by $C_1\langle \delta | f_E(\delta)\rangle + C_{1\delta}\langle \delta | \delta f_E({\delta})\rangle $. The $C_1$ term is easy to calculate: $C_1 p(\hat{k}) P_{\delta}(k)$, where $P_{\delta}$ can be evaluated at one-loop order to be consistent with the other terms. It can also be evaluated at arbitrary precision in the context of this perturbative expansion without introducing unphysical effects (see discussion in \cite{blazek15}). In the following, we choose to use the fully nonlinear $P_{\rm NL}$, for instance using the Halofit perscription \cite{smith03,takahashi12}.

At one-loop order, the $C_{1\delta}$ term can be expanded:
\begin{align}
\langle \delta | \delta f_E(\delta)\rangle &= \left[\langle \delta^{(2)} | \delta^{(1)} f_E(\delta^{(1)})\rangle + \langle \delta^{(1)} | \delta^{(2)} f_E(\delta^{(1)})\rangle + \langle \delta^{(1)} | \delta^{(1)} f_E(\delta^{(2)})\rangle\right] \\
&\equiv A_{0|0E} + B_{0|0E} + C_{0|0E}. \notag
\end{align}

These terms are:
\begin{align}
A_{0|0E}(k,\mu_k) &= 2 \int \frac{d^3q}{(2\pi)^3} f_E(\hat{q}) F_2(\mathbf{q_2},\mathbf{q}) \Pl(q)\Pl(q_2) ~,\\
B_{0|0E}(k,\mu_k) &= \left\{\frac{8}{105}\sigma^2 p(\hat{k}) \Pl(k)\right\} ~,\\
C_{0|0E}(k,\mu_k) &= \left\{\frac{10}{21}\sigma^2 p(\hat{k}) \Pl(k) \right\}\ + 2 \Pl(k) \int \frac{d^3q}{(2\pi)^3} \Pl(q) \left[f_E(\hat{q}_2) F_2(-\mathbf{q},\mathbf{k}) - \frac{5}{21}p(\hat{k})  \right] ~.
\end{align}

Where we define $\mathbf{q_2} = \mathbf{k}-\mathbf{q}$ and $\sigma^{2n}=\int^{\Lambda}\frac{d^3q}{(2\pi)^3} \Pl(q)^n$, which is dependent on the high-$k$ cutoff (e.g.\ the relevant smoothing scale -- see Sec.~\ref{sec:smoothing}). Note that we have explicitly separated the $k\rightarrow 0$ contribution to the integral in $C_{0|0E}$, placing it in brackets. This contribution, along with $B_{0|0E}$ are both proportional to the linear $C_1$ term. As discussed below in Sec.~\ref{sec:renorm}, we renormalize the $C_1$ parameter by absorbing these cutoff-dependent terms. Thus, the terms in brackets are not included in any subsequent evaluations of these functions, and we omit the analogous terms in subsequent expressions. 
In total, the resulting power spectrum is given by:
\begin{align}
\label{eq:ta_dE}
P_{\delta E}(k,\mu_k) = 
   C_1 p(\hat{k}) P_{\delta}(k)
+ C_{1 \delta} \left[A_{0|0E}(k,\mu_k) + C_{0|0E}(k,\mu_k)\right] +\mathcal{O}(P_{L}^3) ~.
\end{align}

We note the similarity between the $C_{0|0E}$ term and the terms that contribute to the third-order non-local galaxy bias $b_{\rm 3, NL}$ \cite{mcdonald09b,saito14}. Both arise from the nonlinear evolution of tidal field and consist of the linear power spectrum multiplied by a filtered power spectrum. We speculate that this term, and similar terms seen below, will combine with the $t_{ij}$ contribution to provide an analogous ``non-local'' contribution to IA. We will explore this issue further in \cite{schmitz_inprep}.

\subsubsection{II correlation}
The II term is given by $C_1^2\langle f_E(\delta) | f_E(\delta)\rangle + 2C_1 C_{1\delta}\langle f_E(\delta) | \delta f_E(\delta)\rangle + C_{1\delta}^2\langle \delta f_E(\delta) | \delta f_E(\delta)\rangle$. As before, the first term is straightforward: $C_1^2 p(\hat{k})^2 P_{\delta}(k)$. The $C_1 C_{1\delta}$ term has the same form as the $C_{1\delta}$ contribution to GI above (although with a different dependence on $\mu_k$ beyond the Limber approximation). The $C_{1\delta}^2$ term has only one correlator at one-loop order. However, this term can contribute both $E$- and $B$-modes. Although pure tidal alignment ($\gamma_{ij} \propto s_{ij}$) produces only $E$-modes (just as gravitational lensing does at leading order), $C_{1\delta}$ corresponds to a modulation of the signal by the density weighting, which converts some of the $E$-modes into $B$-modes. The $E$-modes are given by:
\begin{align}
\langle \delta f_{E}(\delta) | \delta f_{E}(\delta)\rangle &= \langle \delta^{(1)} f_{E}(\delta^{(1)}) | \delta^{(1)} f_{E}(\delta^{(1)})\rangle \\
&\equiv A_{0E|0E} \notag\\
&= \int \frac{d^3q}{(2\pi)^3} \left[\Pl(q)\Pl(q_2)\left(f_E(\hat{q})f_E(\hat{q_2}) + f^2_E(\hat{q})\right) -\frac{8}{15} p(\hat{k})^2 \Pl^2(q)\right] ~, \notag
\end{align}
where the equivalent expression holds for $A_{0B|0B}$, and we have explicitly subtracted out the $k\rightarrow 0$ piece, which is absorbed into the effective ``shape noise'' contribution (along with similar terms below -- see Sec.~\ref{sec:shapenoise}). In total, we have:
\begin{align}
\label{eq:ta_EE}
P_{EE}(k,\mu_k) &= C_1^2 p(\hat{k})^2 P_{\delta}(k) + 2 C_1 C_{1\delta} p(\hat{k}) \left[A_{0|0E}(k,\mu_k)+ C_{0|0E}(k,\mu_k)\right] + C_{1\delta}^2A_{0E|0E}(k,\mu_k) +\mathcal{O}(\Pl^3) ~,\\
\label{eq:ta_BB}
P_{BB}(k,\mu_k) &= C_{1\delta}^2A_{0B|0B}(k,\mu_k) +\mathcal{O}(\Pl^3) ~.
\end{align}
Note that the frequently used (and ambiguously named) ``nonlinear linear alignment model'' (NLA) consists of the first terms in Eqs.~\ref{eq:ta_dE} and \ref{eq:ta_EE}, where a fully nonlinear model is used for $P_{\delta}$.

\subsection{Tidal torquing}

\subsubsection{GI correlation}
We write all potential terms involving $C_{2}$ or $C_{2\delta}$:
\begin{align}
C_2 \langle \delta | h(\delta,\delta)\rangle + C_{2\delta} \langle \delta | \delta h(\delta,\delta)\rangle&= C_2 \left[\langle \delta^{(2)} | h(\delta^{(1)},\delta^{(1)})\rangle
+2 \langle \delta^{(1)} | h(\delta^{(2)},\delta^{(1)})\rangle \right]
+C_{2\delta}\langle \delta^{(1)} | \delta^{(1)}h(\delta^{(1)},\delta^{(1)})\rangle \\
&\equiv C_2 \left[ A_{0|E2} + B_{0|E2} \right] + C_{2\delta} C_{0|0E2}~. \notag
\end{align}
These power spectra are:
\begin{align}
A_{0|E2}(k,\mu_k) &= 2 \int \frac{d^3\mathbf{q}}{(2\pi)^3} \Pl(q) \Pl(q_2)F_2(\mathbf{q},\mathbf{q_2})h_E(\hat{q},\hat{q}_2) ~,\label{ttgia}\\
B_{0|E2}(k,\mu_k) &= 4 \Pl(k) \int \frac{d^3\mathbf{q}}{(2\pi)^3} \Pl(q) \left[F_2(\mathbf{q},\mathbf{-k})h_E(\hat{q},\hat{q}_2)- \frac{29}{630}p(\hat{k})\right] ~,\label{ttgib}\\
C_{0|0E2}(k,\mu_k)&= 0 ~.
\end{align}
We have explicitly removed the $k \rightarrow 0$ contribution to $B_{0|E2}(k)$, which is absorbed (renormalized) into the definition of $C_1$. Note that $A_{0|E2}(k\rightarrow0)=0$. 
\subsubsection{II correlation}
At $\mathcal{O}(\Pl^2)$, only one term contributes, $C_2^2 \langle h(\delta^{(1)},\delta^{(1)})| h(\delta^{(1)},\delta^{(1)})\rangle$, corresponding to:
\begin{align}
\label{ttii0}
P_{EE}(k,\mu_k)&=C_2^2 A_{E2|E2}(k,\mu_k) ~, \\
A_{E2|E2}(k,\mu_k) &= 2\int \frac{d^3\mathbf{q}}{(2\pi)^3} \left[\Pl(q) \Pl(q_2) h^2_{E}(\hat{q},\hat{q}_2) - \frac{4}{135} p(\hat{k})^2 \Pl^2(q)\right] ~,
\end{align}
and the equivalent for $P_{BB}$.
We have subtracted off the constant $k \rightarrow 0$ contribution.

\subsection{Cross-terms in the II correlation}
For galaxy populations where both linear and quadratic alignments are relevant, the II correlation will have a contribution from cross terms:
\begin{align}
&2C_1 C_2 \langle f(\delta) | h(\delta,\delta)\rangle 
+2C_1 C_{2\delta} \langle f(\delta) | \delta h(\delta,\delta)\rangle 
+2C_{1\delta} C_2 \langle \delta f(\delta) | h(\delta,\delta)\rangle \\
&= 2 C_1 C_2 \left[\langle f(\delta^{(2)}) | h(\delta^{(1)},\delta^{(1)})\rangle
+2 \langle f(\delta^{(1)}) | h(\delta^{(2)},\delta^{(1)})\rangle \right]\notag\\
&~~~~~~+2 C_1 C_{2\delta}\langle f(\delta^{(1)}) | \delta^{(1)}h(\delta^{(1)},\delta^{(1)})\rangle
+2 C_{1\delta} C_2\langle \delta^{(1)}f(\delta^{(1)}) | h(\delta^{(1)},\delta^{(1)})\rangle \notag\\
&\equiv 2 C_1 C_2 \left[ A_{E|E2} + B_{E|E2}\right] + 2 C_1 C_{2\delta} C_{E|0E2} + 2 C_{1\delta} C_2 D_{0E|E2} ~. \notag
\end{align}
These power spectra are:
\begin{align}
A_{E|E2}(k,\mu_k) &= p(\hat{k})A_{0|E2} = 2 p(\hat{k})\int \frac{d^3\mathbf{q}}{(2\pi)^3} \Pl(q) \Pl(q_2)F_2(\mathbf{q},\mathbf{q_2})h_E(\hat{q},\hat{q}_2) ~,\label{ttiia}\\
B_{E|E2}(k,\mu_k) &= p(\hat{k})B_{0|E2} = 4 \Pl(k) p(\hat{k}) \int \frac{d^3\mathbf{q}}{(2\pi)^3} \Pl(q) \left[F_2(\mathbf{q},\mathbf{-k})h_E(\hat{q},\hat{q}_2) - \frac{29}{630} p(\hat{k})\right] ~,\label{ttiib}\\
C_{E|0E2}(k,\mu_k) &= 0 ~, \\
D_{0E|E2}(k,\mu_k) &= 2 \int \frac{d^3\mathbf{q}}{(2\pi)^3} \left[ \Pl(q) \Pl(q_2)f_{E}(\hat{q}_2) h_{E}(\hat{q},\hat{q}_2) - \frac{4}{45} p(\hat{k})^2 \Pl^2(q)\right] ~,
\label{ttiid}
\end{align}
with the equivalent expression for $D_{0B|B2}$ (only this term can contribute to $P_{BB}$).
As before, we have subtracted the $k \rightarrow 0$ contribution to the $B$ and $D$ terms.

%%%%%%%%%%%%
\subsection{Complete Model}
\label{sec:complete}

Combining all terms, we obtain the following expressions for the GI and II power spectra:
\begin{align}
\label{eq:dEtot}
P_{\delta E}(k, \mu_k) =& C_1 p(\hat{k}) P_{\delta}(k) + C_{1\delta} \left[A_{0|0E}(k,\mu_k) + C_{0|0E}(k,\mu_k)\right]
+ C_2 \left[ A_{0|E2}(k,\mu_k) + B_{0|E2}(k,\mu_k)\right]~, \\
\label{eq:EEtot}
P_{EE}(k, \mu_k) =&
C_1^2 p(\hat{k})^2 P_{\delta}(k) + 2 C_1 C_{1\delta} p(\hat{k}) \left[A_{0|0E}(k,\mu_k)+ C_{0|0E}(k,\mu_k)\right] + C_{1\delta}^2 A_{0E|0E}(k,\mu_k)\\
&+ C_2^2 A_{E2|E2}(k,\mu_k)
+ 2 C_1 C_2 p(\hat{k}) \left[ A_{0|E2}(k,\mu_k) + B_{0|E2}(k,\mu_k)\right] + 2 C_{1\delta} C_2 D_{0E|E2}(k,\mu_k)~,\notag\\
\label{eq:BBtot}
P_{BB}(k, \mu_k) =& C_{1\delta}^2 A_{0B|0B}(k,\mu_k) + C_2^2 A_{B2|B2}(k,\mu_k) + 2 C_{1\delta} C_2 D_{0B|B2}(k,\mu_k) ~.
\end{align} 

\begin{figure*}[hbt]
\includegraphics[width=\linewidth, trim=0cm 0cm 0cm 0cm, clip=true]{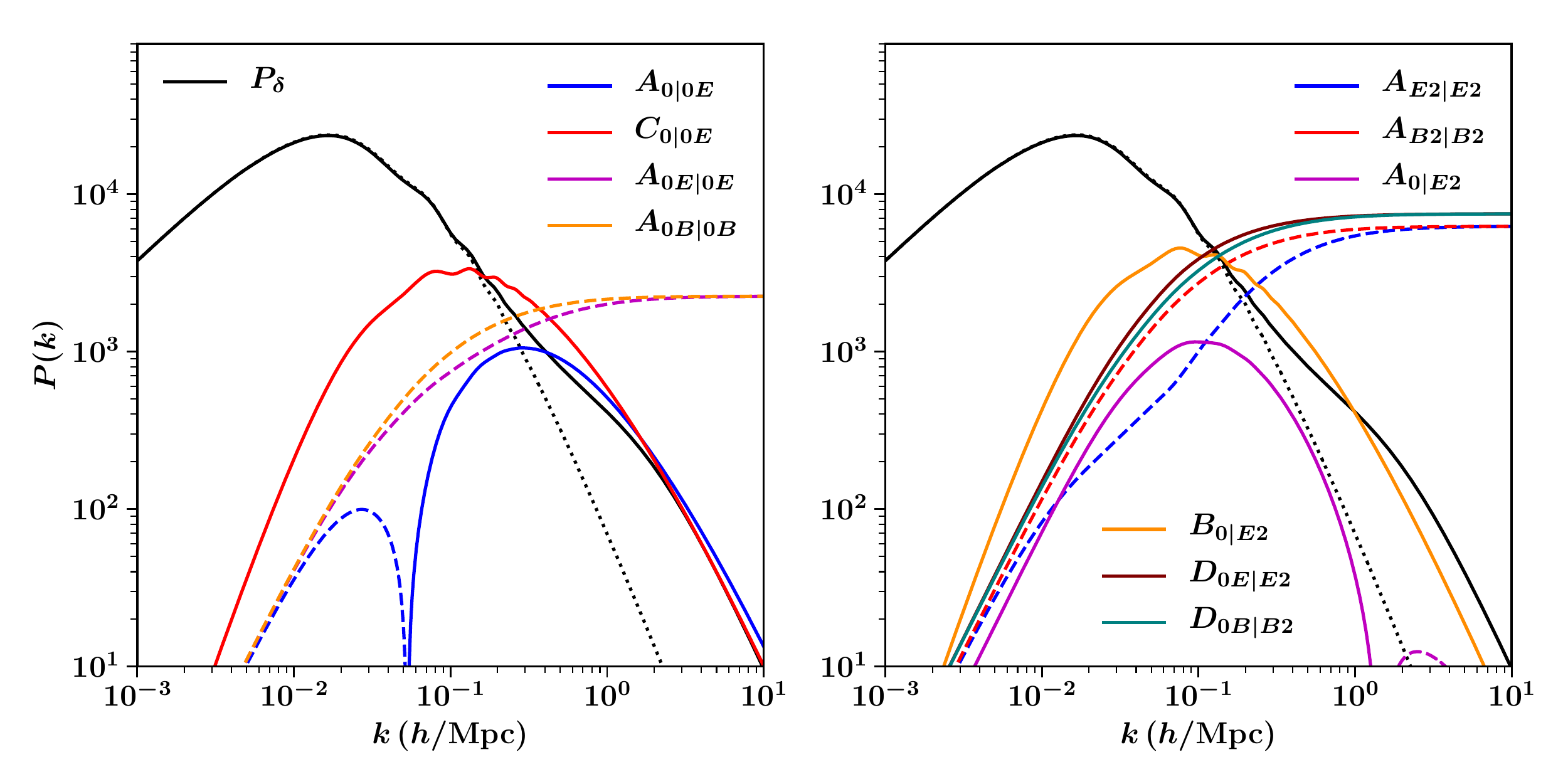}
\caption{The components of the $z=0$, II power spectra from are shown. The pre-factors in Eqs.~\ref{eq:EEtot}-\ref{eq:BBtot} are included, with $C_1 = C_{1\delta} = -1$ and $C_2 = 5$, corresponding to the fiducial relative scaling between $C_1$ and $C_2$, without the factor of $\bar{C_1}\rho_{\rm crit}\Omega_{\rm m}$ in Eqs.~\ref{eq:C1amp}-\ref{eq:C2amp}. We assume transverse modes ($\mu_k = 0$). Negative values are indicated with dashed lines. {\it Left panel}: contributions from tidal alignment ($C_1$ and $C_{1\delta}$). {\it Right panel}: contributions from tidal torquing ($C_2$) and mixed terms. For reference, in both panels the leading tidal alignment contribution $C_1^2 P_{\delta}$ is shown, with the solid line for $P_{\rm NL}$ (the NLA model) and the dotted line for $\Pl$.}
\label{fig:pk}
\end{figure*}

In Fig.~\ref{fig:pk}, we show all model components to the II term. Since the GI term is made of a subset of the II components, we do not show it separately.
In the following subsections, we discuss some of the technical details of the model.
%%%%%%%%%%%%
%%%%%%%%%%%%

\subsection{Normalization}
\label{sec:norm}

The parameters $C_i$ are dimensionless numbers that capture the effective large-scale response for each term. In analogy with galaxy bias parameters, they can be treated as general functions of redshift (and galaxy properties) and measured for a given sample. Historically, these parameters have been rescaled to capture the expected amplitude and redshift evolution. As we intend this work to be useful for implementing models in upcoming weak lensing analyses, we will briefly outline these conventions and describe one reasonable set of choices. While these conventions would not impact a fully general IA analysis (where the amplitude and redshift dependence are allowed complete freedom), in practice, parameterizations typically limit the redshift evolution to follow Eq.~\ref{eq:C1amp} below or a variant thereof (e.g.\ with an additional power law in redshift).

Several conventions exist for this normalization rescaling. Early works \cite{catelan01,mackey02} used the ellipticity variance between individual galaxies to set the scale of the amplitude, assuming that these deterministic IA expansions were responsible for (nearly) the entire observed variance. By assuming that stochasticity from smaller scale physics (i.e.\ terms not captured in this perturbative expansion) is negligible, this approach sets an upper limit on the IA amplitude parameters. An additional issue with such an approach is that the value can be highly dependent on the minimum scale of fluctuations considered when calculating the variance (i.e.\ the relevant smoothing scale, whether it is implicit or explicit) -- see \cite{blazek11,larsen16} for further discussion. To partially avoid these issues, \cite{hirata04} used the variance of galaxies smoothed on large angular windows, analogous to how density fluctuations are normalized using the $\sigma_8$ parameter. The particular measurements used in that work, from the low-redshift SuperCOSMOS survey \cite{brown02}, are not particularly well-matched to modern lensing surveys. However, this normalization convention, formalized in \cite{bridle07}, has become fairly standard, and as recent observations have shown, it provides roughly the correct scale for observed IA correlations (up to an order-unity parameter). 

Note that \cite{hirata04} assumed redshift evolution corresponding to the ``primordial alignment'' scenario, in which the tidal field at high redshift, around galaxy formation, was responsible for the observed IA at late times. Outside of the redshift dependence, the choice of ``primordial'' or ``instantaneous'' alignment can be thought of as the Lagrangian or Eulerian description, respectively. In the context of a complete effective theory at a given order, these two approaches should be equivalent. We note that our current treatment emits one term at $\mathcal{O}(\delta^2)$, the velocity shear $t_{ij}$, and will revisit this topic in upcoming work \cite{schmitz_inprep}.

\subsubsection{Tidal alignment}

Synthesizing these results, the tidal alignment convention has become (see \cite{blazek15} for further discussion):
\begin{align}
\label{eq:C1amp}
C_1(z) = - A_1(z) \left(\bar{C_1}\rho_{\rm crit}\right) \Omega_m G(z)^{-1} ~.
\end{align}
The minus sign enforces the expected behavior that galaxies (and their host halos) will tend to be oriented towards overdense regions rather than tangentially aligned as results from lensing shear. The number $\bar{C_1} = 5\times10^{-14} h^{-2}M_{\odot}^{-1}{\rm Mpc}^3$, corresponding to $\bar{C_1}\rho_{\rm crit} \approx 0.014$, was determined from the windowed ellipticity variance in SuperCOSMOS and assuming the NLA model (i.e.\ the nonlinear matter power spectrum used with the linear IA model; \cite{bridle07}). The growth factor $G(z)$, normalized to unity at $z=0$, is included to cancel the linear growth of the density field and yield a constant amplitude in the primordial alignment scenario.\footnote{The original treatment in \cite{hirata04} normalized the growth function differently, which would lead to a roughly 30\% difference with the current convention. The value of $\bar{C_1}$ quoted here was determined by \cite{bridle07} using our convention for $G(z)$.} The fractional matter density $\Omega_m$ is factored out to reflect the fact that a larger density increases the amplitude of the tidal field, while the combination $A_1\bar{C_1}$ captures the response to the tidal field.

The remaining free parameter $A_1(z)$ is now expected to be an order-unity parameter that describes the particular galaxy samples and captures potential deviations from the assumed redshift dependence. Current constraints on $A_1$ from cosmic shear measurements (e.g.\ \cite{troxel17arxiv}) are consistent with $A_1 \sim 1$ for typical lensing sources (although there is not yet a strong detection), while direct IA measurements with massive elliptical galaxies (e.g.\ \cite{joachimi11,singh15,okumura09b}) find $A_1 \sim 3-10$, depending on the redshift and luminosity.

As discussed above, if $C_{1\delta}$ is assumed to come purely from density weighting effect, at one-loop order it will take the value $C_{1\delta} = b_1 C_1$. More generally, we can define an analogous scaling:
\begin{align}
\label{eq:C1deltaamp}
C_{1\delta}(z) = - A_{1\delta}(z) \left(\bar{C_1}\rho_{\rm crit}\right) \Om G^{-1}(z)~.
\end{align}

\subsubsection{Tidal torquing}

Conventions for setting the expected amplitude of tidal torquing, $C_2$, have typically relied on assuming that this quadratic term is responsible for the full variance between individual galaxy ellipticities (e.g.\ \cite{mackey02, larsen16}). Instead, we propose following a similar procedure as for $C_1$ in Eq.~\ref{eq:C1amp}, namely that the fiducial pre-factor is set by matching to the observed variance in large angular windows. In this case, we can write:
\begin{align}
\label{eq:C2amp}
C_2(z) = A_2(z) \left(\frac{5\bar{C_1}\rho_{\rm crit}}{\Omega_{\rm m,fid}}\right) \Om^2 G(z)^{-2} ~.
\end{align}
We have multiplied the factor $\bar{C_1}\rho_{\rm crit}$ by 5 to account for the approximate difference in windowed variance produced by the different (unnormalized) IA power spectrum in the pure tidal alignment and tidal torquing cases. Under this convention, the $C_1^2$ and $C_2^2$ contributions to the II term produce approximately the same windowed ellipticity variance at $z=0$, for $|A_1| = |A_2|$. This correction factor is mildly dependent on cosmology, but we choose to apply an approximate and cosmology-independent value for simplicity. Because the quadratic term has two powers of the tidal field, there are two factors of both $\Om$ and $G(z)$, although we divide by a fiducial $\Omega_{\rm m,fid}$ to maintain the correct numerical value.\footnote{In practice, without a precise prediction for the shape response to the tidal field, IA contains no usable information on $\Om$, which can be equivalently absorbed into the pre-factor. The recent analysis in \cite{troxel17arxiv} treats both $C_1$ and $C_2$ as scaling linearly with $\Om$.} The overall difference in sign compared to $C_1$ is to maintain the convention that positive $A_i$ corresponds to galaxy shape alignment with overdense regions (i.e.\ a negative GI contribution). In the case of tidal alignment, a positive $A_1$ corresponds to both the theoretical expectation and what has been widely observed in both real galaxies and simulations. However, the expected sign of $A_2$ is less clear, with some hydrodynamic simulations (e.g.\ \cite{chisari16}) finding tangential alignment between the major axes of spiral galaxies and matter overdensities, corresponding to negative $A_2$, and other simulations (e.g.\ \cite{tenneti16,hilbert17}) finding the opposite. There is not yet strong evidence for $A_2$ in galaxy observations, although the recent analysis of \cite{troxel17arxiv} found hints of $A_2 < 0$. In the absence of a strong indication for either sign, we assume a fiducial $A_2 = 1$ in the following forecasts.

While Eq.~\ref{eq:C2amp} sets a scaling for $C_2$ that is consistent with the motivation for the established convention in Eq.~\ref{eq:C1amp}, it does not necessarily correspond to an equivalent level of overall IA contamination. This scaling is determined using windowed ellipticity variance, a measure of the II term. However, the GI term is often the dominant IA contribution, and this term differs significantly between the linear and quadratic IA contributions. Indeed, because there is no leading-order $C_2$ contribution to GI, the overall IA impact from $C_2$ is suppressed compared to a $C_1$ model with equivalent windowed variance.

Finally, we emphasize that these proposed scalings are somewhat arbitrary, although they are useful in establishing a standardized approach to compare results between surveys. However, given the particular assumptions made in these scalings, care must be taken when limiting the allowed redshift dependence of IA in an analysis.

\subsection{Smoothing}
\label{sec:smoothing}

In several earlier works on IA modeling, a smoothing filter was explicitly applied to the tidal field to remove fluctuations below the halo or galaxy scale (e.g.\ \cite{hirata04,blazek11,blazek15,chisari13}). Following the typical treatment in galaxy bias, we instead choose to treat the smoothing of the tidal (and density) fields as an implicit element of the model, considering correlations only on scales much larger than the smoothing scale $(k \ll k_{\rm sm})$ and incorporating contributions to these correlations from small scales into the effective (renormalized) IA bias parameters $C_i$. As discussed in \cite{angulo15,schmidt15}, the effect of non-local contributions, such as smoothing, can be incorporated through higher derivative operators which will scale as powers of $(k/k_{\rm sm})^2$ and thus become significant on small scales (including where a perturbative expansion will begin to break down). The inclusion of such terms can be similarly motivated by considering a Taylor expansion of a Fourier-space Gaussian smoothing filter. We do not include such terms here but note that they are generically present and may reflect the small-scale physics of galaxy and halo formation relevant to IA correlations. Accounting for these terms will be especially important when attempting to extend a perturbative expansion to smaller scales and including the impact of the one-halo term.

\subsection{Renormalized contributions}
\label{sec:renorm}
 \subsubsection{IA bias parameters}

As seen above, we absorb the cutoff-dependent contributions (i.e.\ those proportional to $\sigma^2$ or $\sigma^4$) into the definitions of the effective IA parameters. This process is identical to the renormalization of bias parameters (e.g.\ \cite{mcdonald06,mcdonald09b,desjacques16arxiv}), which was inspired by the renormalization of coupling constants in quantum field theory. The underlying principle is that contributions to large-scale correlations sourced by small scale physics can be absorbed into effective bias coefficients which are the observable quantities, rather than the ``bare'' parameters. Because these small-scale processes are not accurately modeled (or modeled at all) in a perturbative expansion, we cannot attempt to predict the amplitude of these contributions through these calculations, even if the integrals are not actually divergent in a $\Lambda$CDM universe. Instead, the resulting parameters are determined through observation, simulation, or more detailed modeling of small-scale physics. Our treatment of these cutoff-dependent contributions in this work is in contrast to \cite{blazek15}, where the preferred approach was to include them as physical contributions determined by the tidal field smoothing. We note that this renormalized parameter approach to IA can be incorporated into a more comprehensive effective field theory (EFT) approach to modeling large scale structure (e.g.\ \cite{senatore15,angulo15}).

Because correlations involving higher-order terms are absorbed into the lower-order parameters, we see that each of these parameters is naturally ``generated'' by higher-order corrections, unless an underlying symmetry forces it to zero. For instance, if we had started with $C_1 = 0$ (e.g.\ a pure tidal torquing scenario), the existence of $C_2 \neq 0$ generates an effective amplitude for $C_1$ (as seen from the $k \rightarrow 0$ limit in the $B(k)$ terms above). Thus all galaxies should exhibit some tidal alignment behavior on sufficiently large scales, even if the underlying astrophysical processes that determine intrinsic shapes are based on tidal torquing, or some other nonlinear process. Although phrased in different language, the generation of this linear term is the same as the contribution to linear galaxy alignment discussed in \cite{hui08} (see also \cite{larsen16}).

\subsubsection{Stochasticity and shape noise}
\label{sec:shapenoise}

We have not explicitly included stochastic contributions into our IA expansion (Eq.~\ref{eq:IAexp}). However, the existence of a random component of galaxy shapes, ``shape noise,'' has long been understood as a critical element to weak lensing measurements because this component is significantly larger than the induced gravitational shears \cite{bernstein02}. Shape noise is typically thought of as an underlying property of the galaxy sample, characterized by one-point ellipticity variance $\sigma^2_{\rm SN}$. As the number density of galaxy shapes is increased, the relative importance of shape noise decreases. In the measured ellipticity power spectrum, if every galaxy shape were measured perfectly and completely uncorrelated, we would expect the resulting shape noise contribution to scale as $\sigma^2_{\rm SN}/n$, for galaxy number density $n$. However, because shapes are imperfectly measured due to measurement noise, an ``effective number density'' is introduced to capture the observed contribution to the ellipticity power spectrum: $\sigma^2_{\rm SN}/n_{\rm eff}$ \cite{amara08,chang13,jarvis16}. Although not typically considered in the same language, correlations between galaxy shapes, i.e.\ intrinsic alignments, will also alter $n_{\rm eff}$. This effect is intuitively straightforward: a correlated ensemble of measured shapes is not sampling the underlying shape distribution as completely as the raw number of objects would suggest.

As before, the similarity with galaxy biasing is strong. In the case of galaxy bias, the leading stochastic contribution is ``shot noise,'' and the basic model is a Poissonian contribution to the power spectrum equal to $n^{-1}$. In perturbative bias expansions, it was seen that quadratic bias $b_2$ and other higher-order terms contributed correlations with a constant $k\rightarrow 0$ limit \cite{mcdonald06,mcdonald09b}. In the spirit of renormalization, to restore the expected (and measured) linear biasing behavior on large-scales, these constant contributions were absorbed into an effective stochasticity term, which was then allowed to display non-Poissonian behavior, including cross-correlations between different samples. Studies of halos in N-body simulations showed that non-Poissonian stochasticity was indeed present \cite{hamaus10}, with large halos displaying sub-Poissonian stochasticity. Models for this behavior, including nonlinear biasing and halo exclusion, have been developed \cite{baldauf13,ginzburg17arxiv}. Importantly, this work has demonstrated that the stochastic component is not expected to be the same in the $k\rightarrow 0$ and $k \rightarrow \infty$ limits. Even if Poissonian shot noise is recovered at $k \rightarrow \infty$, on finite scales, we expect to see non-Poissonian stochasticity, including scale dependence.

Returning to the question of shape noise, it is clear that higher-order terms in the perturbative IA expansion contribute to the observed shape correlations, since some II terms have non-zero $k \rightarrow 0$ limits. We have absorbed these constant contributions into an effective shape noise, which becomes a free parameter, leaving the remaining scale dependence as part of the IA model. Conceptually, this effective shape noise can be expressed through the use of $n_{\rm eff}$, defined to include the absorbed IA contributions. If these contributions induce positive shape correlations, we would expect $n_{\rm eff}$ to decrease, thus increasing shape-noise (analogous to nonlinear galaxy clustering and super-Poissonian shot noise). Conversely, negative shape correlations would increase $n_{\rm eff}$ and reduce shape noise (analogous to halo exclusion and sub-Poissonian shot noise).

However, the central lesson from the galaxy biasing case is that the IA ``shape noise'' contribution to the power spectrum on finite scales is decoupled from the zero-lag shape variance. Thus, while $n_{\rm eff}$ can, in principle, be defined to include IA contributions to shape noise, the method used to estimate it must also include these contributions. Typically, $n_{\rm eff}$ is measured from the one-point ellipticity variance, appropriately weighted by measurement noise (e.g.\ \cite{jarvis16}), and will not include these IA effects. Moreover, the IA contributions can impact the ellipticity cross-power spectrum between different samples, leading to a free shape noise, $\epsilon_{ij}$, for correlations between redshift bins $i$ and $j$. In practice, estimating the ``true'' $\sigma^2_{\rm SN}/n_{\rm eff}$ from the data is equivalent to measuring $\epsilon_{ij}$ along with the lensing and IA signals. There may be additional information in higher-point correlations, which we leave for future work. If galaxy shapes are dominated by physics on sub-halo scales, then shape noise on measurable scales will likely still be reasonably well described by the traditional approach based on the zero-lag shape variance. In this case, the zero-lag estimate may be sufficient for determining the covariance and would provide fairly tight priors on $\epsilon_{ij}$ for Fourier space analyses. In configuration space, a free constant shape noise only impacts the covariance (since changing the power spectrum by an overall constant only affects the correlation function at zero-lag).

Finally, we note that there is often no meaningful distinction between ``stochasticity'' and physics on scales below what is modeled. As biasing and IA theory becomes more sophisticated, including a fully nonlinear treatment of the one-halo term, much of what is currently called ``stochasticity'' will become deterministic features of the model.

%%%%%%%%%%%%%%%%
\begin{figure*}[hbt]
\includegraphics[width=\linewidth, trim=0cm 0cm 0cm 0cm, clip=true]{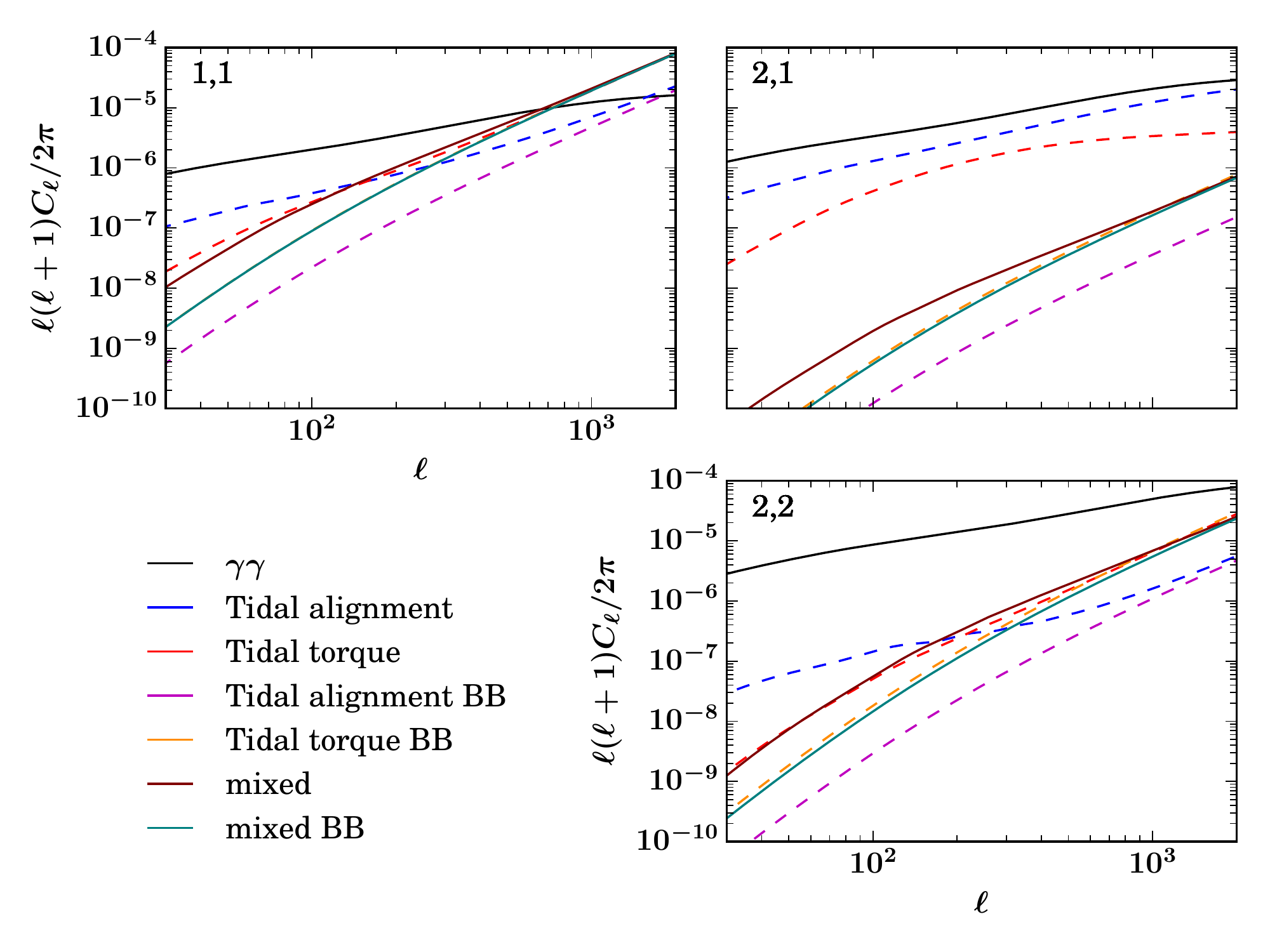}
\caption{IA contributions to the angular auto- and cross-power spectra for two source bins with Gaussian $n(z)$, with means $\langle z \rangle=0.4$ and 0.8 and width $\Delta z=0.1$. Dashed lines indicate negative values, and $B$-modes are denoted ``BB.'' For reference, the lensing contribution is shown in black.}
\label{fig:cls}
\end{figure*}
%%%%%%%%%%%%%%%%

%%%%%%%%%%%%%%%%
\section{Model implementation and impact}
\label{sec:implementation}
%%%%%%%%%%%%%%%%
\subsection{Implementation}
We have implemented the integrals in Sec.~\ref{sec:corr} using the {\tt FAST-PT} code \cite{mcewen16,fang17}, which uses FFTs to decompose the input linear power spectrum into power-law components for which the convolution integrals can be analytically performed, allowing for extremely rapid evaluation (see also \cite{schmittfull16}). These capabilities are included in the public versions\footnote{Available at https://github.com/JoeMcEwen/FAST-PT.} of {\tt FAST-PT}, beginning with v2.1. Appendix~\ref{app:fastpt} shows the decomposition of each term into the basis of {\tt FAST-PT} \cite{fang17}.

We have also incorporated the {\tt FAST-PT} code as a module in the cosmological inference package {\tt CosmoSIS} \cite{zuntz15}, which we use to perform the forecasts below. Fig.~\ref{fig:cls} shows the contributions to the observed angular shape power spectra, $C_{\ell}$, from this implementation of our model, assuming two equal number density source bins with Gaussian $n(z)$, with mean redshifts of 0.4 and 0.8 and width $\Delta z=0.1$. A similar {\tt CosmoSIS} implementation was used in the recent DES Year 1 tomographic cosmic shear analysis \cite{troxel17arxiv}. That analysis assumed that $C_{1 \delta} = C_1$, corresponding to the case where the $C_{1 \delta}$ term arises purely from density weighting of the lensing sources with linear bias $b_1 = 1$.
%%%%%%%%%%%%
\subsection{Impact on cosmic shear constraints from a Stage IV weak lensing survey}\label{sec:impact}

Future wide-field galaxy imaging surveys such as LSST, Euclid, and WFIRST will use cosmic shear to constrain several cosmological parameters, including the dark energy equation of state. A tomographic (i.e.\ multiple redshift bin) analysis allows the separation of the observed correlations into those from weak lensing shear and those due to IA. In this section, we implement a simulated likelihood analysis to assess the biases in inferred cosmological parameters when assuming the wrong intrinsic alignment model. Operationally, we use a theoretical prediction for the observed correlations (including IA contamination when relevant) and use this ``data vector'' as input to a parameter estimation analysis, along with a suitable covariance matrix. An analysis combining galaxy clustering and weak lensing (including the cross-correlation between the two, i.e.\ galaxy-galaxy lensing) can add additional constraints on intrinsic alignments as well as calibration of other systematic errors such as photometric redshift biases (e.g.\ \cite{joachimi10,bernstein09,des_keypaper_17arxiv}).
While the models presented here are applicable to such a combined probes analysis, we choose to limit this impact study to cosmic shear for simplicity. We assume that the $E$-mode angular power spectra $C_{\ell}$ are the two-point statistic used, and therefore we do not include the $B$-mode contributions. Note that a real-space analysis using shear correlation functions, which are a mixture of $E$- and $B$-modes, would need to include these contributions. 

We assume an $18000\ \mathrm{deg}^2$ LSST-like survey with an effective number density of source galaxies of 30 ${\rm arcmin}^{-2}$, with redshift distribution parameterized as \cite{smail95}
\begin{align}
P(z) = z^{\alpha} \exp\left[-\frac{z}{z_0}^\beta\right] ~,
\end{align}
with $\alpha=1.23$, $z_0=0.51$ and $\beta=1.01$, following \cite{chang13}. 
We assume that the photometric redshift estimate used to place galaxies in redshift bins is Gaussian distributed around the true redshift with scatter $\sigma(z)=0.05(1+z)$. Finally, we assume the sample is divided into 5 equal number density redshift bins, and use for our data vector the angular shear power spectra, $C_{\ell}$ for all redshift bin combinations in the multipole range $100<l<1000$. We use the Gaussian approximation for the covariance (e.g.\ \cite{hu04}) -- we assume the shear field is a Gaussian random field, and do not include the effects of a realistic survey geometry or super-sample covariance (see e.g.\ \cite{takada13}). As discussed in Sec.~\ref{sec:shapenoise}, we do not assume to know the constant contribution to the observed $C_{\ell}$ due to shape noise; we marginalize over a free constant contribution for each redshift bin pair.\footnote{We do this marginalization analytically by adding a large number to each block diagonal in the covariance matrix. We have verified that this approach is equivalent to explicitly marginalizing over a free additive parameter for each redshift bin with prior range $-1\times 10^{-3}$ to $1\times10^3$.}
Apart from intrinsic alignments, we do not include systematics nuisance parameters in this forecast (e.g.\ the photo-$z$ distributions are fixed to their correct values). We thus do not consider this a fully realistic forecast of the constraining power of LSST, but rather an instructive demonstration of the impact of this more sophisticated IA model and the potential systematic bias in the inferred cosmological parameters due to IA. 

For the likelihood analyses we vary $\Om$ and $\sig$ and the dark energy equation of state using the two-parameter model \cite{chevallier01,linder03}
\begin{align}
w(a) = w_0 + w_a(1-a)~,
\end{align}
where $a$ is the scale factor of the Universe normalized to unity at the present. We use uniform priors $0.1<\Om<0.6$, $0.5<\sig<1.1$, $-3<w_0<-0.3$, and $-3<w_a<3$.

We perform the following likelihood analyses:
\begin{enumerate}[(i)]
\item{We generate a fake data vector without intrinsic alignment contributions ($A_1=A_2=0$) and do not include intrinsic alignments in the modeling. This should trivially produce unbiased constraints on the cosmological parameters and provide a baseline for the statistical constraining power under our set of assumptions.}
\item{We generate a fake data vector with the full intrinsic alignment model and fiducial amplitudes ($A_{1} = A_{2} = 1$), but use the NLA model in the analysis, with a uniform prior on the single amplitude of $-5<A_1<5$. This case serves to test the impact of using the NLA model (see e.g.\ \cite{heymans13,DES15,hildebrandt17} for recent cosmic shear analyses that assumed this model) if the galaxy alignments are in fact described by the full model.
We expect the inferred cosmological parameters to be biased to some extent in this case, since the intrinsic alignment model used is insufficiently flexible to describe the data vector.}
\item{We repeat (ii), but add an additional free parameter to the NLA model -- a power-law in redshift such that the NLA amplitude is $A_1(z) = (1+z)^\alpha A_1(0)$ (see e.g.\ \cite{maccrann15,joudaki17} for recent cosmic shear analyses that used this model).}
\item{We use the same fake data vector as in case (ii), but now use the full intrinsic alignment model in the likelihood analysis, marginalizing over $A_1$ and $A_2$ in the range $[-5,5]$. This case should also produce unbiased constraints.}
\end{enumerate}

\begin{figure}[h!]
\centering
\includegraphics[width=\columnwidth, trim=1cm 1cm 1cm 1cm, clip=true]{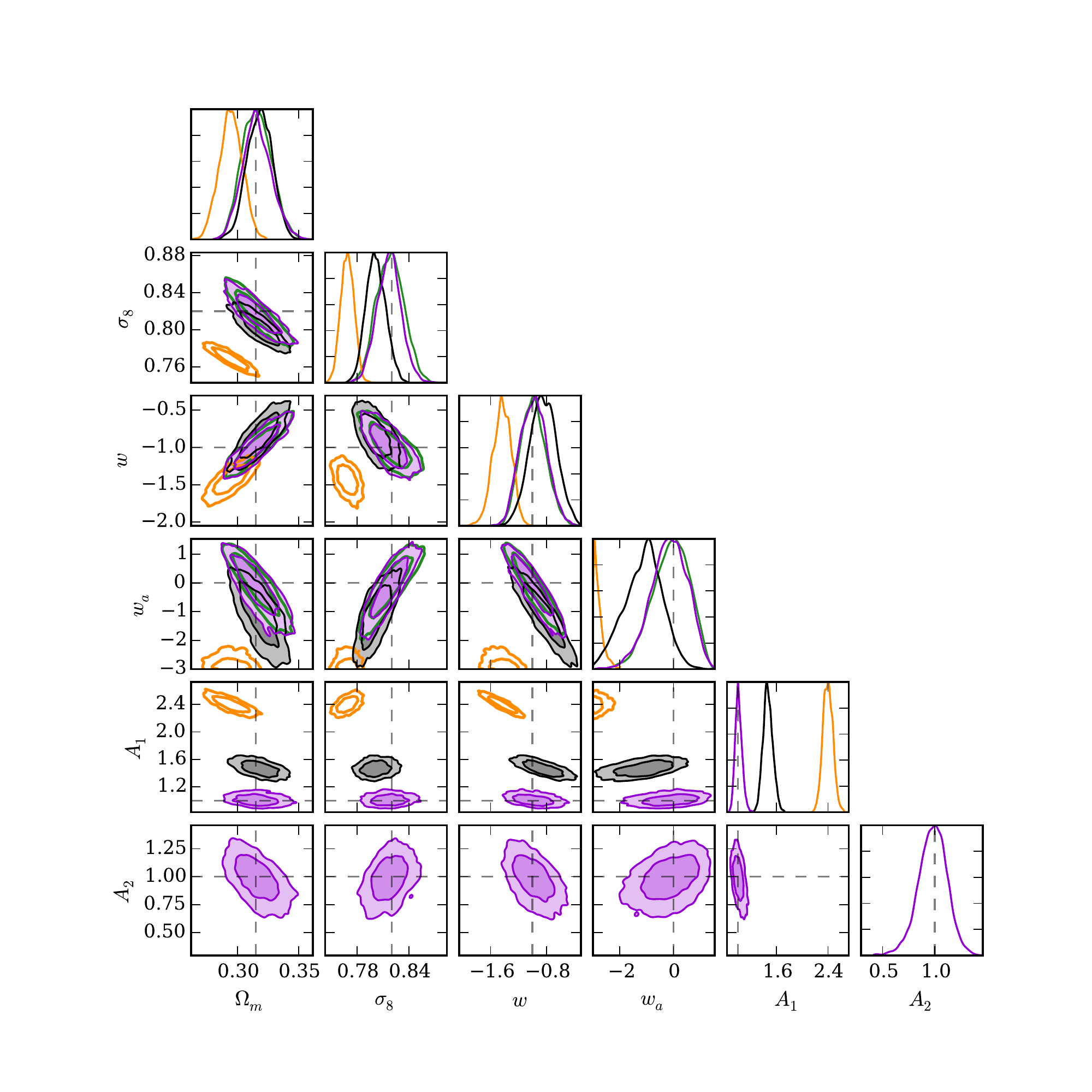}
\caption[]{Constraints on cosmological and intrinsic alignment parameters for an idealized LSST-like cosmic shear survey. Dashed lines indicate the input parameter values used to create the data vectors. Green outlined contours use a data vector and model without intrinsic alignment contributions, case (i). The orange outlined contour uses a data vector with contamination by the full intrinsic alignment model, with fiducial amplitudes (see Sec.~\ref{sec:impact}), but uses a model which assumes the NLA model for the intrinsic alignment contribution, case (ii). The black contour is the same as the orange, except the model also includes a free power law in redshift, case (iii). The purple contour uses the same data vector as orange and black, but uses the full intrinsic alignment modeling, thereby recovering unbiased parameter constraints, case (iv).}
\label{fig:corner}
\end{figure}

Figure~\fig{fig:corner} shows the results of these forecasts. The green, unfilled contours are the $68\%$ and $95\%$ parameter credible intervals for case (i): no intrinsic alignment contamination in the data vector or the model. It is instructive to consider the constraints on $w(a_{\mathrm{piv}})$, where $a_{\mathrm{piv}}$ is the scale factor at which $w(a)$ is best constrained. We find $a_{\mathrm{piv}}=0.75$ ($z=0.33$) and use this value in all of the following quoted constraints. Given that we generate our data vectors assuming a \lcdm\ universe, the true value of $w(a_{\mathrm{piv}})$ is $-1$. % (i.e. we do not update $a_{\mathrm{piv}}$ for each of the following set of constraints).
For this case, we find $w(a_\mathrm{piv})=-1.01\pm0.06$ (68\%).

The orange unfilled contour represents case (ii): the data vector is contaminated by the full intrinsic alignment model with fiducial amplitudes, but is modeled assuming the NLA model. Clearly this scenario results in large biases in most parameters, for example $w(a_\mathrm{piv})=-2.13 \pm 0.13$. The grey filled contour corresponds to case (iii): an extra free parameter is varied, allowing the NLA amplitude to vary with redshift according to a power-law. Because the linear and quadratic terms can have significantly different redshift dependence (e.g.\ when assuming the scaling described in Sec.~\ref{sec:norm}), this additional freedom in the model should allow for a more accurate fit. Indeed, this case performs significantly better, however it still results in biased inferred parameters, with $w(a_\mathrm{piv})=-1.13 \pm 0.13$ -- the systematic bias from assuming the wrong IA model is roughly equivalent to the statistical uncertainty.  Finally, the purple filled contour represents case (iv): the full IA model is used to generate the data vector and is used in the parameter estimation, with the amplitudes $A_1$ and $A_2$ marginalized over. As expected, the correct cosmological parameters are recovered. We find $w(a_\mathrm{piv})=-1.03 \pm 0.08$. The uncertainty on the inferred value of $w(a_\mathrm{piv})$ is increased by $2\%$ compared to the case when no intrinsic alignment parameters are marginalized over.

Marginalization over a free shape noise parameter for each auto- and cross-spectrum removes some constraining power, since there is non-trivial degeneracy between such an additive term and the overall amplitude of the lensing signal. This marginalization produces constraints that are qualitatively similar to those from a correlation function analysis, where this constant term appears only at zero-lag and is thus inaccessible, although a more detailed quantitative comparison is challenging due to the inherent mixing of scales when transforming between configuration and Fourier space (see \cite{lee17inprep} for a more detailed discussion in the context of galaxy clustering). Even in the absence of the IA contributions discussed in this work, this type of marginalization is likely necessary for future Fourier-space cosmic shear analyses, given the statistical precision of the measurements and the uncertainties in estimating the shape noise. Finally, we note that the constraints on some cosmological parameters can actually be slightly improved by the presence of IA, even if the amplitude parameters are unknown, since there is cosmological information in the shape of the IA power spectrum (e.g.\ the constraints on $\Omega_m$ and $\sigma_8$ in Fig.~\ref{fig:corner}). Accessing this information requires sufficiently good photo-$z$ information to separate the IA and lensing signals through tomography -- as this information decreases, the degeneracy between the two will increase, and the overall cosmological constraints will degrade in the presence of IA. Alternatively, combining weak lensing with galaxy clustering and galaxy-galaxy lensing will enable better separation of these signals. We leave more detailed consideration of these analyses for future work.

%%%%%%%%%%%%
%%%%%%%%%%%%

%%%%%%%%%%%%%%%%
\section{Conclusions}
\label{sec:conc}
%%%%%%%%%%%%%%%%

Given the current uncertainty in the IA of typical lensing sources and its potentially significant impact, it is important that future lensing experiments use sufficiently sophisticated modeling. We have presented a perturbative model for intrinsic alignments, motivated by the treatment of galaxy biasing, which incorporates both tidal alignment and tidal torquing mechanisms and allows for more general alignment effects. We have also performed forecasts that show the potential impact of IA if an insufficient model is adopted. Using the traditional NLA model when the true underlying IA signal has quadratic contributions leads to systematic biases in the inferred cosmology significantly larger than the underlying statistical uncertainty. These biases partially remain even when allowing for a more flexible redshift dependence in the NLA model.

The recent weak lensing analysis of DES Year 1 data \cite{troxel17arxiv} applied this model and found indications for non-zero values of both $C_1$ and $C_2$, respectively at the 82\% and 84\% confidence levels. Using this more flexible IA model caused a non-trivial shift in the recovered cosmological parameters, although they caution that further study is required to understand this result. This model will also provide a valuable tool in ``combined probe'' analyses that use both weak lensing and galaxy clustering information to improve statistical information and break degeneracies between both cosmological and astrophysical/systematics parameters (e.g.\ \cite{des_keypaper_17arxiv}). By including correlations between the (biased) density field and the intrinsic shapes, such analyses allow a more effective separation of the IA and lensing signals and thus better measurement of both. However, optimal combined probe analyses will require consistent, nonlinear modeling of IA, galaxy biasing, and their cross-correlation. The perturbative approach described here provides exactly such a description.

Cosmological hydrodynamic simulations also present an opportunity to test the predictions of this model. However, the current state of IA measurements in these simulations has not converged, with fairly low signal-to-noise (driven by the maximum volumes that can be simulated) and a strong dependence on sub-grid physics which leads to a lack of qualitative and quantitative agreement between simulations. We expect that the interaction of analytic theory and hydrodynamic simulations will be a valuable area of study in the near future.

Finally, this modeling approach reveals interesting theoretical features of IA. Due to IA parameter renormalization, we see that tidal (linear) alignment terms are generated even when starting with a higher-order astrophysical model (e.g.\ tidal torquing). More generally, this renormalization approach provides a framework in which the small-scale physics of galaxy formation and evolution are responsible for the shape correlations observed on large scales, thus motivating a significant dependence of the IA parameters on galaxy properties. Similarly, the nonlinear IA correlations include $k\rightarrow 0$ contributions, suggesting that the effective shape noise contribution to measured shear correlations may be decoupled from the zero-lag ellipticity variance and should instead be treated as a free parameter of the model. More generally, exploring the connection between IA and galaxy biasing is yielding valuable insights, and we believe it will continue to do so as the modeling of both further develops in the new era of lensing measurements.

\acknowledgments
We thank Christopher Hirata, Mike Jarvis, Benjamin Joachimi, Fabian Schmidt, Denise Schmitz, Zvonimir Vlah, and David Weinberg for helpful discussions. We also thank Sarah Bridle for hosting a workshop at the University of Manchester, where much of this work was done. JB is supported by an SNSF Ambizione Fellowship. XF is supported by the Simons Foundation.

%\bibliographystyle{apsrev4-1}
%\bibliography{refs_local}


%%%%%%%%%%%%
\pagebreak
\appendix
%%%%%%%%%
%\onecolumngrid
%%%%%%%%%
\section{Angular kernels}
\label{app:kernels}
%%%%%%%%%%%%%%%%

The evaluation of the convolution integrals is simplified if we rotate to a spherical coordinate system in which $\hat{k}$ is the polar axis, allowing us to analytically integrate over the azimuthal angle. For the alternative {\tt FAST-PT} decomposition, see Appendix~\ref{app:fastpt}. We express the resulting integrals in terms of $\alpha = q/k$ and $\mu = \hat{k}\cdot\hat{q}$. For convenience, we define the following functions:
\begin{align}
X_1 &=\alpha ^2-2 \alpha  \mu +1~,\\
X_2 &=-3 \left(\alpha ^2+1\right) \mu ^2+\alpha ^2+3 \alpha  \mu ^3+\alpha  \mu +1~,\\
X_3 &=\alpha  \left(10 \mu ^2-3\right)-7 \mu~,\\
X_4 &= -10 \alpha + 7 (1 + \alpha^2) \mu - 4 \alpha \mu^2~,\\
X_5 &=\alpha ^3 \left(-38 \mu ^5+4 \mu ^3+2 \mu \right)+\alpha^2  \left(19 \mu ^6+44 \mu ^4-17 \mu^2+2\right)+ \alpha \left( -34 \mu ^5-4 \mu ^3+6 \mu \right)~,\\
X_6 &=19 \mu ^4-14 \mu ^2+3~,\\
X_7 &=\left(1-\mu ^2\right) (\alpha -\mu )^2 (1-2\alpha  \mu)^2~,\\
X_8 &=-4+12\mu^2+4\alpha\mu(1-9\mu^2) +\alpha^2 (-3+10\mu^2+41\mu^4) + \alpha^3\mu(9-22\mu^2-19\mu^4) + \alpha^4 X_6 ~,\\
X_9 &= -2 + 4 \alpha \mu - \alpha^2 (-1 + 3 \mu^2)~,\\
X_{10} &=-1 - 2 \mu^2 + 19 \mu^4 + \alpha^2 (6 - 28 \mu^2 + 38 \mu^4) + \alpha (2 \mu + 4 \mu^3 - 38 \mu^5)~,\\
X_{11} &= 6 + 12 \mu^2 - 50 \mu^4 - 4 \alpha^2 (1 - 18 \mu^2 + 25 \mu^4) + 4 \alpha \mu (-7 - 2 \mu^2 + 25 \mu^4) ~,\\
X_{12} &=  -1 - 18 \mu^2 + 35 \mu^4 + 2 \alpha \mu (9 + 10 \mu^2 - 35 \mu^4) + \alpha^2 (6 - 60 \mu^2 + 70 \mu^4)~,\\
X_{13} &= (-1 + \mu^2) (-1 + 5 \mu^2)~,\\
X_{14} &= \alpha (-1 + \mu^2) (-4 \mu + \alpha (-1 + 5 \mu^2))~, \\
X_{15} &= 2 \mu \left(\mu^2-1\right) \left( \frac{\alpha-\alpha^2\mu}{X_1}-\mu \right)~,\\
X_{16} &= 1 - 12 \mu^2 + 19 \mu^4 + 2 \alpha^3 \mu (7 + 2 \mu^2 - 25 \mu^4) + \alpha^4 (1 - 18 \mu^2 + 25 \mu^4) \\
&+ \alpha (6 \mu + 8 \mu^3 - 
    46 \mu^5) + \alpha^2 (4 - 41 \mu^2 + 60 \mu^4 + 25 \mu^6)~, \notag\\
X_{17} &= 3 - 18 \mu^2 + 23 \mu^4 + 2 \alpha^3 \mu (9 + 10 \mu^2 - 35 \mu^4) + 2 \alpha \mu (7 + 2 \mu^2 - 25 \mu^4) \notag\\
&+ \alpha^4 (3 - 30 \mu^2 + 35 \mu^4) + \alpha^2 (-2 - 45 \mu^2 + 60 \mu^4 + 35 \mu^6)~, \\
X_{18} &= \left(\mu^2-1\right)(-1 + 2 \mu^2 - 2 \alpha \mu (2 + \mu^2) + \alpha^4 (-1 + 5 \mu^2) - 2 \alpha^3 (\mu + 5 \mu^3) + \alpha^2 (3 + 
    5 (\mu^2 + \mu^4))) ~,\\
X_{19} &= -2 (-4 + 12 \mu^2 + 8 \alpha \mu (1 - 5 \mu^2) + \alpha^4 (1 - 18 \mu^2 + 25 \mu^4) + \alpha^2 (-5 - 2 \mu^2 + 
      55 \mu^4) \\
      &+ \alpha^3 \mu (23 - 5 \mu^2 (6 + 5 \mu^2))) ~,\notag\\
X_{20} &= -4 + 12 \mu^2 + 4 \alpha \mu (3 - 11 \mu^2) + \alpha^4 (3 - 30 \mu^2 + 35 \mu^4) + \alpha^2 (-11 - 6 \mu^2 + 
    65 \mu^4) \\
    &+ \alpha^3 \mu (33 - 5 \mu^2 (6 + 7 \mu^2)) ~,\notag\\
X_{21} &= 2\alpha(\mu-\alpha)(\alpha\mu-1)(2\alpha\mu-1)(\mu^2-1) ~,\\
X_{22} &= \alpha (1-\mu^2) (2 \mu + \alpha (-3 + \alpha^2 + 5 \alpha \mu - 5 (1 + \alpha^2) \mu^2 + 5 \alpha \mu^3))
 ~.
\end{align}
We can then express the relevant correlations:
\begin{align}
A_{0|0E}(k, \mu_k)&= 2 (1-\mu_k^2) \int \frac{k^3\alpha^2d\alpha d\mu}{(2\pi)^2}  \frac{(3\mu^2-1)X_3}{28\alpha X_1} \Pl(q) \Pl(q_2) ~, \\
C_{0|0E}(k, \mu_k)&= 2 (1-\mu_k^2) \Pl(k)\int \frac{k^3\alpha^2d\alpha d\mu}{(2\pi)^2} 
\left( \frac{X_4 X_{9}}{28\alpha X_1} -\frac{5}{21} \right) \Pl(q) ~, \\
A_{0E|0E}(k, \mu_k)&= \int \frac{k^3\alpha^2d\alpha d\mu}{(2\pi)^2}  \left[ \left(\frac{X_{10}+ X_{11}\mu_k^2 + X_{12}\mu_k^4}{8X_1}\right)\Pl(q) \Pl(q_2)-\frac{8}{15}\Pl^2(q)\right]~, \\
A_{0B|0B}(k, \mu_k)&= \int \frac{k^3\alpha^2d\alpha d\mu}{(2\pi)^2}   \left[ \left( X_{15}
+ \left( \frac{X_{13}}{2} + \frac{X_{14}}{2X_1}\right)\mu_k^2 \right)
\Pl(q) \Pl(q_2)-\frac{8}{15}\Pl^2(q)\right]~, \\
A_{0|E2}(k, \mu_k) &= 2(1-\mu_k^2)\int \frac{k^3\alpha^2d\alpha d\mu}{(2\pi)^2}  \frac{X_2 X_3}{84X_1^2 \alpha} \Pl(q) \Pl(q_2) ~, \\
B_{0|E2}(k, \mu_k) &= 4(1-\mu_k^2)\Pl(k) \int \frac{k^3\alpha^2d\alpha d\mu}{(2\pi)^2}  \left( \frac{X_2 X_4}{84X_1 \alpha} - \frac{29}{630}\right) \Pl(q) ~,\\
A_{E2|E2}(k, \mu_k) &=2\int \frac{k^3\alpha^2d\alpha d\mu}{(2\pi)^2} \left[
\frac{X_5 + (1+\alpha^4)X_6 - 2 X_{16} \mu_k^2 + X_{17}\mu_k^4}{72 X_1^2} \Pl(q) \Pl(q_2) 
 - \frac{4}{135}\Pl^2(q)\right]~, \\
A_{B2|B2}(k, \mu_k) &=2\int \frac{k^3\alpha^2d\alpha d\mu}{(2\pi)^2} \left[ \frac{X_7 + X_{18}\mu_k^2}{18 X_1^2} \Pl(q) \Pl(q_2)
- \frac{4}{135}\Pl^2(q)\right]~, \\
D_{0E|E2}(k, \mu_k) &= 2\int \frac{k^3\alpha^2d\alpha d\mu}{(2\pi)^2} \left[\frac{X_8 + X_{19} \mu_k^2 + X_20\mu_k^4}{24 X_1^2} \Pl(q) \Pl(q_2)
 - \frac{4}{45}\Pl^2(q)\right]~, \\
D_{0B|B2}(k, \mu_k) &= 2\int \frac{k^3\alpha^2d\alpha d\mu}{(2\pi)^2} \left[\frac{X_{21}+X_{22}\mu_k^2}{6 X_1^2}  \Pl(q) \Pl(q_2)
 - \frac{4}{45}\Pl^2(q)\right]~.
\end{align}
The remaining angular integral for $B_{0|E2}$ and $C_{0|0E}$ can be be done analytically, yielding:
\begin{align}
B_{0|E2}(k) &= 4 (1-\mu_k^2) \Pl(k) k^3 \int \frac{\alpha^2 d\alpha}{(2\pi)^2} \Pl(q) \left[ Z_1(\alpha) - \frac{29}{315}\right] ~,\\
C_{0|0E}(k) &= 2 (1-\mu_k^2) \Pl(k) k^3 \int \frac{\alpha^2 d\alpha}{(2\pi)^2} \Pl(q) \left[ Z_2(\alpha) - \frac{10}{21}\right] ~,\\
Z_1(\alpha) &= \frac{2 \alpha (225 - 600 \alpha^2 + 1198 \alpha^4 - 600 \alpha^6 + 
   225 \alpha^8) + 225 ( \alpha^2-1)^4 (\alpha^2+1) \log \left|\frac{\alpha-1}{\alpha+1}\right| }{20160 \alpha^5} ~,\\
Z_2(\alpha) &= \frac{4 \alpha (45 + 379 \alpha^2 - 165 \alpha^4 + 45 \alpha^6) + 
 90 (\alpha^2-1)^4\log \left|\frac{\alpha-1}{\alpha+1}\right| }{1344 \alpha^3} ~.
 \end{align}
For large $\alpha$ (relevant at low $k$), the leading behavior of the kernels simplifies:
\begin{align}
\alpha^2\left[Z_1(\alpha)-\frac{29}{315}\right] \rightarrow  -\frac{16}{147} + \frac{353}{4704\alpha^2} + \mathcal{O}(\alpha^{-4}) ~, \\
\alpha^2\left[Z_2(\alpha)-\frac{10}{21}\right] \rightarrow  \frac{24}{49} - \frac{31}{784 \alpha^2} + \mathcal{O}(\alpha^{-4}) ~.
\end{align}

%%%%%%%%%%%%%%%%
\section{Integrals in {\tt FAST-PT}}
\label{app:fastpt}
%%%%%%%%%%%%%%%%

The evaluation of these integrals can performed significantly faster by decomposing the convolution kernels into terms with different dependences on wavevector amplitude and angle (expressed as an expansion in Legendre polynomials). The angular part of the resulting component integrals can be performed analytically, while the remaining 1d convolution over wavenumber can be performed rapidly using FFTs. Below, we provide the decomposition of the relevant terms into this basis, described in \cite{fang17} and implemented in the publicly available code {\tt FAST-PT}. Note that the current implementation assumes the Limber approximation ($\mu_k=0$).

As discussed above, the $k \rightarrow 0$ contributions are removed from each term and absorbed into either the renormalized IA parameters or the effective shape noise. In the {\tt FAST-PT} decomposition, $\propto \Pl$ contributions of this type are removed through kernel redefinition. However, the shape noise terms, $\propto \sigma^4$, cannot be removed from the convolution before evaluation and must be explicitly subtracted from the result. {\tt FAST-PT} returns $\sigma^4$ for the input power spectrum, which can be used to perform this subtraction from the final results, but it is not done by default to allow the user to control numerical precision.

Below, we list the terms that are output by the relevant {\tt FAST-PT} functions, denoted with the {\tt F} superscript, related to the integrals in Sec.~\ref{sec:corr}. In cases where there is an overall pre-factor in front of the integral, it is applied after combining the individual components, and it is thus not included in the {\tt FAST-PT} coefficients quoted here. $C_{0|0E}$ and $B_{0|E2}$ are $P_{13}$-like integrals. The analytic forms for these terms in App.~\ref{app:kernels} are directly implemented in {\tt FAST-PT} using discrete convolutions \cite{mcewen16}.

\begin{align}
\label{eq:0|0E}
A^{\tt F}_{0|0E} = 2 \int\frac{d^3\mathbf{q}}{(2\pi)^3}f_E(\hat{q})F_2(\mathbf{q_2},\mathbf{q})\Pl(q)\Pl(q_2) ~.
\end{align}
The corresponding coefficients shown in Table \ref{tab:IAdeltaE1}.

\begin{table}[hbt]
\centering
\begin{tabular}{|c| c | c | c | c | c |}
 \hline
 \rule{0pt}{2.5ex} $\alpha$ & $\beta$ & $\ell$ & $\ell_1$ & $\ell_2$ & $A_{\ell_1\ell_2\ell}^{\alpha\beta}$\\
 \hline
  $0$&$0$& $0$& $0$ & $2$ & $\nicefrac{17}{21}$ \\
     &   & $2$& $0$ & $2$ & $\nicefrac{4}{21}$ \\
 \hline
 $1$& $-1$ &$1$& $0$ & $2$ & $\nicefrac{1}{2}$ \\
 \hline
 $-1$& $1$ &$1$& $0$ & $2$ & $\nicefrac{1}{2}$ \\
 \hline
\end{tabular}
\caption{The coefficient of each term in the Legendre polynomial expansion of the $\left[f_E(\hat{q})F_2(\mathbf{q_2},\mathbf{q})\right]$ kernel in Eq.~\ref{eq:0|0E}.}
\label{tab:IAdeltaE1}
\end{table}

\begin{align}
A^{\tt F}_{0E|0E}= \int\frac{d^3\mathbf{q}}{(2\pi)^3}\left[f_E(\hat{q})f_E(\hat{q}_2)+f_E(-\hat{q})f_E(\hat{q})\right]\Pl(q)\Pl(q_2)~,
\label{eq:0E0E}
\end{align}
with $A^{\tt F}_{0B|0B}$ given by $E\rightarrow B$.
The coefficients are in Table \ref{tab:0E0E}, where $\alpha=\beta=0$ for all the terms.

\begin{table}[hbt]
\centering
\begin{tabular}{| c | c | c | c | c |}
 \hline
 \rule{0pt}{2.5ex} $\ell$ & $\ell_1$ & $\ell_2$ & $A_{\ell_1\ell_2\ell}^{00(E)}$ & $A_{\ell_1\ell_2\ell}^{00(B)}$\\
 \hline
 $0$& $0$ & $0$ & $\nicefrac{29}{90}$ & $\nicefrac{2}{45}$\\
       &  & $4$ & $\nicefrac{19}{35}$ & $-\nicefrac{16}{35}$\\
       & $2$ & $0$ & $\nicefrac{5}{63}$ & $-\nicefrac{44}{63}$ \\
       &     & $2$ & $\nicefrac{19}{18}$ & $-\nicefrac{8}{9}$ \\
 \hline
 $1$& $1$ & $1$ & --- & $2$  \\
 \hline
\end{tabular}
\caption{The coefficient of each term in the Legendre polynomial expansion of $\left[f_{(E,B)}(\hat{q}_1)f_{(E,B)}(\hat{q}_2)+f_{(E,B)}(-\hat{q}_1)f_{(E,B)}(\hat{q}_1)\right]$ in Eq.~\ref{eq:0E0E}.}
\label{tab:0E0E}
\end{table}

\begin{align}
A^{\tt F}_{E2|E2}(k) = 2\int \frac{d^3\mathbf{q}}{(2\pi)^3} \left[\Pl(q) \Pl(q_2) h^2_{E}(\hat{q},\hat{q}_2)\right]~,
\label{eq:E2E2}
\end{align}
with $A^{\tt F}_{B2|B2}$ given by $E\rightarrow B$.
The coefficients are in Table \ref{tab:E2E2}, where $\alpha=\beta=0$ for all the terms.

\begin{table}
\centering
\begin{tabular}{| c | c | c | c | c |}
 \hline
 \rule{0pt}{2.5ex} $\ell$ & $\ell_1$ & $\ell_2$ & $A_{\ell_1\ell_2\ell}^{00(E)}$ & $A_{\ell_1\ell_2\ell}^{00(B)}$\\
 \hline
 $0$& $0$ & $0$ & $\nicefrac{16}{81}$ & $-\nicefrac{41}{405}$\\
       & $2$ & $0$ & $\nicefrac{713}{1134}$ & $-\nicefrac{298}{567}$ \\
       & $2$ & $2$ & $\nicefrac{95}{162}$ & $-\nicefrac{40}{81}$ \\
       & $4$ & $0$ & $\nicefrac{38}{315}$ & $-\nicefrac{32}{315}$\\
 \hline
 $1$& $1$ & $1$ & $-\nicefrac{107}{60}$ & $\nicefrac{59}{45}$  \\
       & $3$ & $1$ & $-\nicefrac{19}{15}$ & $\nicefrac{16}{15}$  \\
 \hline
 $2$& $0$ & $0$ & $\nicefrac{239}{756}$ & $-\nicefrac{2}{9}$ \\
       & $2$ & $0$ & $\nicefrac{11}{9}$ & $-\nicefrac{20}{27}$ \\
       & $2$ & $2$ & $\nicefrac{19}{27}$ & $-\nicefrac{16}{27}$ \\
 \hline
 $3$& $1$ & $1$ & $-\nicefrac{7}{10}$ & $\nicefrac{2}{5}$ \\
 \hline
 $4$& $0$ & $0$ & $\nicefrac{3}{35}$ & --- \\
 \hline
\end{tabular}
\caption{The coefficient of each term in the Legendre polynomial expansion of the $h_E^2$ and $h_B^2$ kernels. Due to symmetry under $\ell_1\leftrightarrow \ell_2$, we need only keep terms with $\ell_1\geq \ell_2$ (and have multiplied the coefficients by two where relevant).}
\label{tab:E2E2}
\end{table}

\begin{align}
A^{\tt F}_{0|E2}(k) &= 2 \int \frac{d^3\mathbf{q}}{(2\pi)^3} \Pl(q) \Pl(q_2)F_2(\mathbf{q},\mathbf{q_2})h_E(\hat{q},\hat{q}_2)~.
\end{align}
The coefficients are shown in Table \ref{tab:IA_A}.

\begin{table}[hbt]
\centering
\begin{tabular}{|c| c | c | c | c | c |}
 \hline
 \rule{0pt}{2.5ex} $\alpha$ & $\beta$ & $\ell$ & $\ell_1$ & $\ell_2$ & $A_{\ell_1\ell_2\ell}^{\alpha\beta}$\\
 \hline
  $0$&$0$& $0$& $0$ & $0$ & $-\nicefrac{31}{210}$ \\
     &   & $0$& $2$ & $0$ & $-\nicefrac{34}{63}$ \\
     &   & $2$& $0$ & $0$ & $-\nicefrac{47}{147}$ \\
     &   & $2$& $2$ & $0$ & $-\nicefrac{8}{63}$ \\
     &   & $1$& $1$ & $1$ & $\nicefrac{93}{70}$ \\
     &   & $3$& $1$ & $1$ & $\nicefrac{6}{35}$ \\
     &   & $4$& $0$ & $0$ & $-\nicefrac{8}{245}$ \\
 \hline
 $1$& $-1$ &$1$& $0$ & $0$ & $-\nicefrac{3}{10}$ \\
    &      &$1$& $2$ & $0$ & $-\nicefrac{1}{3}$ \\
    &      &$0$& $1$ & $1$ & $\nicefrac{1}{2}$ \\
    &      &$2$& $1$ & $1$ & $1$ \\
    &      &$1$& $0$ & $2$ & $-\nicefrac{1}{3}$ \\
    &      &$3$& $0$ & $0$ & $-\nicefrac{1}{5}$ \\
 \hline
\end{tabular}
\caption{The coefficient of each term in the Legendre polynomial expansion of the $\left[F_2(\mathbf{q},\mathbf{q_2})h_E(\hat{q},\hat{q}_2)\right]$ kernel.}
\label{tab:IA_A}
\end{table}

\begin{align}
\label{eq:0E|E2}
D^{\tt F}_{0E|E2}(k) &= 2 \int \frac{d^3\mathbf{q}}{(2\pi)^3} \left[ f_{E}(\hat{q}_2) h_{E}(\hat{q},\hat{q}_2) \right]\Pl(q) \Pl(q_2) ~,
\end{align}
and the equivalent for $D^{\tt F}_{0B|B2}$. The coefficients are given in Table \ref{tab:D_EEBB}, with $\alpha=\beta=0$ for all terms.

\begin{table}[hbt]
\centering
\begin{tabular}{| c | c | c | c | c |}
 \hline
 \rule{0pt}{2.5ex} $\ell$ & $\ell_1$ & $\ell_2$ & $A_{\ell_1\ell_2\ell}^{00(E)}$ & $A_{\ell_1\ell_2\ell}^{00(B)}$\\
 \hline
 $0$& $0$ & $0$ & $-\nicefrac{43}{540}$ & $\nicefrac{13}{135}$\\
       &  & $2$ & $-\nicefrac{167}{756}$ & $\nicefrac{86}{189}$\\
       &  & $4$ & $-\nicefrac{19}{105}$ & $\nicefrac{16}{105}$\\
       &$2$& $2$ & $-\nicefrac{19}{54}$ & $\nicefrac{8}{27}$\\
 \hline
 $2$& $0$ & $0$ & $\nicefrac{1}{18}$ & $\nicefrac{2}{9}$\\
       & $2$ & $0$ & $-\nicefrac{7}{18}$ & $\nicefrac{4}{9}$ \\
 \hline
 $1$& $1$ & $1$ & $\nicefrac{11}{20}$& $-\nicefrac{13}{15}$  \\
    &     & $3$ & $\nicefrac{19}{20}$& $-\nicefrac{4}{5}$  \\
 \hline
\end{tabular}
\caption{The coefficient of each term in the Legendre polynomial expansion of $ \left[ f_{(E,B)}(\hat{q}_2) h_{(E,B)}(\hat{q},\hat{q}_2) \right]$ in Eq.~(\ref{eq:0E|E2}).}
\label{tab:D_EEBB}
\end{table}
\end{document}